\documentclass[12pt]{article}
\usepackage{xcolor, graphicx}
\usepackage{subfigure}
\usepackage{amsmath, amssymb}
\usepackage{mathrsfs}
\usepackage{caption}
\usepackage[figuresright]{rotating}
\usepackage{longtable}
\usepackage{geometry}
\usepackage{amsfonts}
\usepackage{color}
\usepackage{tabularx}
\usepackage{multirow}
\usepackage{multicol}
\usepackage{array}
\usepackage{cite}
\usepackage{booktabs}
\usepackage{rotating}
\usepackage{tikz}
\usepackage{epstopdf}
\usepackage[toc,page]{appendix}

\allowdisplaybreaks[4]

\renewcommand{\theequation}{\arabic{section}.\arabic{equation}}

\renewcommand{\Re}{\operatorname{Re}}
\renewcommand{\Im}{\operatorname{Im}}
\renewcommand{\theequation}{\arabic{section}.\arabic{equation}}

\def\sI{\textrm{sI}}

\def\maxI{\textrm{maxI}}
\def\minI{\textrm{minI}}

\def\V={{{\bf\rm{V}}}}

\def\beq{\begin{equation}}
\def\eeq{\end{equation}}
\def\bea{\begin{eqnarray}}
\def\eea{\end{eqnarray}}
\def\ba{\begin{array}}
\def\ea{\end{array}}
\def\no{\nonumber}

\title{Surface energy and elementary excitations of the XYZ spin chain with integrable open boundary fields}
\author{Zhirong Xin${}^{a}$, Junpeng Cao$^{b,c,d,e}$ \thanks{Corresponding author. E-mail: junpengcao@iphy.ac.cn}, Wen-Li Yang$^{d,f,g,h}$\thanks{Corresponding author. E-mail: wlyang@nwu.edu.cn}, and Yupeng Wang$^{b,d}$}

\begin{document}
\date{}
\maketitle
\begin{center}
${}^a$ School of Physics and Electronic Information, Baicheng Normal University, China\\
${}^b$ Beijing National Laboratory for Condensed Matter Physics, Institute of Physics, Chinese Academy of Sciences, Beijing 100190, China\\
${}^c$ School of Physical Sciences, University of Chinese Academy of Sciences, Beijing, China\\
${}^d$ Peng Huanwu Center for Fundamental Theory, Xian 710127, China\\
${}^e$ Songshan Lake Materials Laboratory, Dongguan, Guangdong 523808, China\\
${}^f$ Institute of Modern Physics, Northwest University, Xi'an 710127, China\\
${}^g$ School of Physics, Northwest University, Xi'an, 710127, China\\
${}^h$ Shaanxi Key Laboratory for Theoretical Physics Frontiers, Xi'an 710027, China
\end{center}

\begin{abstract}
We study the thermodynamic limit of the anisotropic XYZ spin chain with non-diagonal integrable open boundary conditions. Although the $U(1)$-symmetry is broken,
by using the new parametrization scheme, we exactly obtain the surface energy and the excitation energy of the system, which has solved the difficulty in the inhomogeneous $T-Q$ relation.
With the boundary parameters in the regions making the Hamiltonian Hermitian, we have obtained the distribution patterns of the zero roots of the eigenvalue of the transfer matrix for the
ground state and the excited ones. We find that the surface and excitation energies depend on the parities of sites number $N$, due to the long-range Neel order in the bulk.
The spontaneous magnetization and easy-axis for all the regions of boundary parameters are studied. We also obtain the physical quantities in the thermodynamic limit of boundary XXZ model by taking the triangular limit.

\vspace{0.5truecm}

\end{abstract}

\section{Introduction}

The anisotropic XYZ quantum spin chain is a typical $U(1)$ symmetry breaking quantum integrable system and has many applications in the statistical mechanics, quantum magnetism, string theory and mathematical physics \cite{baxter2016exactly,korepin1993quantum,vsamaj2013introduction}.
The exact solution of the XYZ model with periodic boundary condition was obtained by Baxter \cite{baxter1971eight,baxter1971one,baxter1972partition,baxter1972one} based on the inversion relations, where the eigenvalues of the transfer matrix
is characterized by the $T-Q$ relation. Later, Takhtadzhan and Faddeev \cite{takhtadzhan1979quantum} resolved the model and proposed the famous quantum inverse scattering method or algebraic Bethe ansatz \cite{korepin1993quantum,sklyanin1978anharmonic}.
We note that all these approaches require that the number of sites is even. Based on the resulting exact solutions, many interesting physical quantities such as elementary excitations and free energy at finite temperature in the
thermodynamic limit have been calculated. Please see \cite{takahashi1999thermodynamics} and references therein.

The next problem is how to obtain the exact solution of the XYZ spin chain with odd number of sites or with the non-diagonal boundary reflections. Although it has been proved that in these cases the model is integrable, it is very hard to get the exact solution, because
the algebraic Bethe ansatz does not work due to the lacking of reference state. Many efforts have be done in this directions and many interesting methods are proposed, such as the separation of variables \cite{57,58},
face-vertex correspondence \cite{60}, $q$-Ansager algebra \cite{3,4,5,6}, modified algebraic Bethe ansatz \cite{13,14,15,16}, and off-diagonal Bethe ansatz \cite{cao2013off,cao2014spin,cao2013off-xyz,wang2015off}.
The most important result is that the eigenvalue of the system is given by the inhomogeneous $T-Q$ relation.

Another problem arises. The inhomogeneous $T-Q$ relation induces the inhomogeneous Bethe ansatz equations. Then the traditional thermodynamic Bethe ansatz may be  very hard to
approach the thermodynamic limit. However, the thermodynamic limit is very important because the number of sites in the real system must be infinite.
We can not ignore this problem.

In this paper, we study the thermodynamic limit of the XYZ model with general integrable open boundary conditions,
where the reflection matrices are non-diagonal and have six free parameters.
The main idea is as follows. We characterize the eigenvalues of the transfer matrix, which is an elliptic polynomial, by its zero points instead of Bethe roots \cite{qiao2020exact,qiao2021exact}.
We obtain the homogeneous equations for the zero points. By the numerical calculation and singularity analysis, we obtain the patterns
of the zero roots at the ground state. Then we calculate the distribution of roots, ground state energy density and surface energy.
This scheme can also be used to study the properties of the system at the thermal equilibrium state.

The paper is organized as follows.
In the next section, the model Hamiltonian and exact solution are introduced.
We study the thermodynamic limit of the general integrable open XYZ model with real crossing parameter $\eta$ in section 3
and with imaginary crossing parameter $\eta$ in section 4.
As an important component, we consider the degenerate case and obtain the results of anisotropic XXZ model in section 5.
Section 6 is the concluding remarks and discussions.
Some useful identities about the elliptic functions are listed in Appendix.

\section{The XYZ spin chain and its exact solutions }
\setcounter{equation}{0}

\subsection{The Hamiltonian}

The Hamiltonian of the XYZ spin chain with generic boundary condition reads
\begin{eqnarray}\label{Hamil}
   H &=& \sum^{N-1}_{j=1} \left( J_x \sigma^x_{j}\sigma^x_{j+1}+J_y \sigma^y_{j}\sigma^y_{j+1}+J_z \sigma^z_{j}\sigma^z_{j+1} \right) +h^-_x\sigma^x_1 +h^-_y\sigma^y_1 +h^-_z\sigma^z_1 \nonumber\\
   && +h^+_x \sigma^x_N +h^+_y\sigma^y_N +h^+_z\sigma^z_N,
\end{eqnarray}
where $\sigma^{\alpha}_j~(\alpha=x,y,z)$ is the Pauli matrix on the $j$th site along the $\alpha$ direction and $N$ is the number of sites.
The anisotropic couplings in the bulk are
\begin{equation}\label{jxyz}
  J_x=\frac{ e^{i\pi \eta}\sigma (\eta+\frac{\tau}{2})}{\sigma(\frac{\tau}{2})},\quad J_y=\frac{ e^{i\pi \eta}\sigma (\eta+\frac{1+\tau}{2})}{\sigma(\frac{1+\tau}{2})},\quad J_z=\frac{ \sigma (\eta+\frac{1}{2})}{\sigma(\frac{1}{2})},
\end{equation}
where $\eta$ is the crossing parameter and $\tau$ is the modulus parameter.
The boundary magnetic fields are
\begin{eqnarray}
  h^{\mp}_x &=& \pm e^{-i\pi(\sum^3_{l=1} \alpha^{\mp}_l-\frac{\tau}{2} )} \frac{\sigma(\eta)}{\sigma(\frac{\tau}{2})} \prod^3_{l=1} \frac{\sigma(\alpha^\mp_l-\frac{\tau}{2})}{\sigma(\alpha^\mp_l)} , \label{hfx}\\
  h^{\mp}_y &=& \pm e^{-i\pi(\sum^3_{l=1} \alpha^{\mp}_l-\frac{1+\tau}{2} )} \frac{\sigma(\eta)}{\sigma(\frac{1+\tau}{2})} \prod^3_{l=1} \frac{\sigma(\alpha^\mp_l-\frac{1+\tau}{2})}{\sigma(\alpha^\mp_l)} ,\label{hfy}\\
  h^{\mp}_z &=& \pm  \frac{\sigma(\eta)}{\sigma(\frac{1}{2})} \prod^3_{l=1} \frac{\sigma(\alpha^\mp_l-\frac{1}{2})}{\sigma(\alpha^\mp_l)},  \label{hfz}
\end{eqnarray}
where $\sigma(u)$ is the $\sigma$-function defined by (\ref{zetaf}) and $\{ \alpha^{\mp}_l | l=1,2,3\}$ are free boundary parameters which specify the strengths of boundary magnetic fields.

The integrability of the model (\ref{Hamil}) is associated with the well-known eight-vertex $R$-matrix $R(u)\in {\rm End}(\mathbb{C}^2\otimes \mathbb{C}^2)$
\begin{equation}\label{Rm}
  R(u )=\left(
    \begin{array}{cccc}
      a(u) &      &      & d(u) \\
           & b(u) & c(u) &  \\
           & c(u) & b(u) &  \\
      d(u) &      &      & a(u) \\
    \end{array}
  \right).
\end{equation}
The non-vanishing matrix entries are \cite{baxter2016exactly}
\begin{eqnarray}
&&\hspace{-2.0truecm} a(u)\hspace{-0.1truecm}=
 \hspace{-0.1truecm}\frac{\theta\left[\begin{array}{c} 0\\\frac{1}{2}
 \end{array}\right]\hspace{-0.16truecm}(u,2\tau)\hspace{0.12truecm}
 \theta\left[\begin{array}{c} \frac{1}{2}\\[2pt]\frac{1}{2}
 \end{array}\right]\hspace{-0.16truecm}(u+\eta,2\tau)}{\theta\left[\begin{array}{c} 0\\\frac{1}{2}
 \end{array}\right]\hspace{-0.16truecm}(0,2\tau)\hspace{0.12truecm}
 \theta\left[\begin{array}{c} \frac{1}{2}\\[2pt]\frac{1}{2}
 \end{array}\right]\hspace{-0.16truecm}(\eta,2\tau)},\quad
b(u)\hspace{-0.1truecm}=\hspace{-0.1truecm}\frac{\theta\left[\begin{array}{c}
 \frac{1}{2}\\[2pt]\frac{1}{2}
 \end{array}\right]\hspace{-0.16truecm}(u,2\tau)\hspace{0.12truecm}
 \theta\left[\begin{array}{c} 0\\\frac{1}{2}
 \end{array}\right]\hspace{-0.16truecm}(u+\eta,2\tau)}
 {\theta\left[\begin{array}{c} 0\\\frac{1}{2}
 \end{array}\right]\hspace{-0.16truecm}(0,2\tau)\hspace{0.12truecm}
 \theta\left[\begin{array}{c} \frac{1}{2}\\[2pt]\frac{1}{2}
 \end{array}\right]\hspace{-0.16truecm}(\eta,2\tau)},\no\\[6pt]
&&\hspace{-2.0truecm}c(u)\hspace{-0.1truecm}=
 \hspace{-0.1truecm}\frac{\theta\left[\begin{array}{c} 0\\\frac{1}{2}
 \end{array}\right]\hspace{-0.16truecm}(u,2\tau)\hspace{0.12truecm}
 \theta\left[\begin{array}{c} 0\\\frac{1}{2}
 \end{array}\right]\hspace{-0.16truecm}(u+\eta,2\tau)}
 {\theta\left[\begin{array}{c} 0\\\frac{1}{2}
 \end{array}\right]\hspace{-0.16truecm}(0,2\tau)\hspace{0.12truecm}
 \theta\left[\begin{array}{c} 0\\\frac{1}{2}
 \end{array}\right]\hspace{-0.16truecm}(\eta,2\tau)},\quad
d(u)\hspace{-0.1truecm}=\hspace{-0.1truecm}\frac{\theta\left[\begin{array}{c}
 \frac{1}{2}\\[2pt]\frac{1}{2}
 \end{array}\right]\hspace{-0.16truecm}(u,2\tau)\hspace{0.12truecm}
 \theta\left[\begin{array}{c} \frac{1}{2}\\[2pt]\frac{1}{2}
 \end{array}\right]\hspace{-0.16truecm}(u+\eta,2\tau)}
 {\theta\left[\begin{array}{c} 0\\\frac{1}{2}
 \end{array}\right]\hspace{-0.16truecm}(0,2\tau)\hspace{0.12truecm}
 \theta\left[\begin{array}{c} 0\\\frac{1}{2}
 \end{array}\right]\hspace{-0.16truecm}(\eta,2\tau)}.\label{Rmd}
\end{eqnarray}
Here  $u$ is the spectral parameter. The associated elliptic functions are defined in Appendix A. In addition to satisfying the quantum Yang-Baxter equation (QYBE)
\begin{eqnarray}
  && R_{12}(u_1-u_2) R_{13}(u_1-u_3) R_{23}(u_2-u_3)  \nonumber \\
  && =  R_{23}(u_2-u_3) R_{13}(u_1-u_3) R_{12}(u_1-u_2), \label{YBE}
\end{eqnarray}
the $R$-matrix also possesses the following properties
\begin{eqnarray}
&& \textrm{Initial conditon:} \quad R_{12}(0)=P_{12}, \\
&& \textrm{Unitarity relation:} \quad R_{12}(u)R_{21}(-u)=-\xi(u) \textrm{id},\quad \xi(u) =\frac{\sigma(u-\eta)\sigma(u+\eta)}{\sigma(\eta)\sigma(\eta)} ,\\
&& \textrm{Crossing relation:} \quad R_{12}(u)=V_{1} R^{t_2}_{12}(-u-\eta)V_{1}, \quad V=-i\sigma^{y} ,\\
&& \textrm{PT-symmetry:} \quad R_{12}(u)=R_{21}(u)=R^{t_1 t_2}_{12}(u), \\
&& \textrm{Z$_2$-symmetry:} \quad \sigma^i_1 \sigma^i_2 R_{12}(u)=R_{12}(u)\sigma^i_1\sigma^i_2,\quad \mbox{for} \quad i=x,y,z, \label{Z2-sym}\\
&& \textrm{Antisymmetry:} \quad R_{12}(-\eta)=-(1-P_{12}) = -2 P^{(-)}_{12}.
\end{eqnarray}
Here $R_{21}(u)=P_{12}R_{12}(u)P_{12}$ with $P_{12}$ being the usual permutation operator and $t_{i}$ denotes transposition in the $i$-th space.
Throughout this paper we adopt the standard notations:
for any matrix $A\in {\rm End}(\mathbb{C}^2 )$, $A_j$ is an embedding
operator in the tensor space $\mathbb{C}^2 \otimes \mathbb{C}^2 \otimes\cdots$,
which acts as $A$ on the $j$-th space and as identity on the other
factor spaces; $R_{i\,j}(u)$ is an embedding operator
in the tensor space, which acts as identity on the factor spaces
except for the $i$-th and $j$-th ones.

Integrable open boundaries can be constructed as follows\cite{sklyanin1988boundary}.
Let us introduce a pair of reflection matrices $K^{-}(u)$ and $K^{+}(u)$.
The former satisfies the reflection equation (RE)
\begin{eqnarray}
  && R_{12}(u_1-u_2)K^{-}_1(u_1)R_{21}(u_1+u_2)K^{-}_2(u_2)  \nonumber \\
  && =  K^{-}_2(u_2) R_{12}(u_1+u_2) K^{-}_1(u_1) R_{21}(u_1-u_2), \label{RE}
\end{eqnarray}
and the latter satisfies the dual RE
\begin{eqnarray}
  && R_{12}(u_2-u_1)K^{+}_1(u_1)R_{21}(-u_1-u_2-2)K^{+}_2(u_2)  \nonumber \\
  && =  K^{+}_2(u_2) R_{12}(-u_1-u_2-2) K^{+}_1(u_1) R_{21}(u_2-u_1) . \label{dual RE}
\end{eqnarray}
Then the transfer matrix $t(u)$ of the open XYZ chain is given by
\begin{equation} \label{tra}
  t(u)=tr_0\{ K^{+}_0(u) T_0(u) K^{-}_0(u) \hat{T}_0(u) \},
\end{equation}
where $T_0(u)$ and $\hat{T}_0(u)$ are the monodromy matrices
\begin{equation}
  T_0(u) = R_{0N}(u-\theta_N) \cdots R_{01}(u-\theta_1), \quad \hat{T}_0(u) = R_{10}(u+\theta_1) \cdots R_{N0}(u+\theta_N).
\end{equation}
Here $\{\theta_j|j=1,\cdots,N  \}$ are the free complex parameters which are usually called as inhomogeneity parameters.

In this paper, we consider the most general solutions of the RE (\ref{RE}) and dual one (\ref{dual RE}), where $K^{\mp}(u)$\cite{inami1994integrable,hou1995solution} read
\begin{eqnarray}
  K^{-}(u) &=& \frac{\sigma(2u)}{2\sigma(u)} \left\{ \textrm{id} + \frac{c^{-}_x \sigma(u)e^{-i\pi u}}{\sigma(u+\frac{\tau}{2})} \sigma^x + \frac{c^{-}_y \sigma(u)e^{-i\pi u}}{\sigma(u+\frac{1+ \tau}{2})} \sigma^y  + \frac{c^{-}_z \sigma(u)}{\sigma(u+\frac{1}{2})} \sigma^z  \right\} , \label{Kmm}\\
  K^+ (u) &=&  K^{-}(-u-\eta)|_{c^{-}_l\rightarrow c^{+}_l} , \label{Kpm}
\end{eqnarray}
where the constants $\{ c ^{\mp}_\alpha| \alpha=x,y,z \}$ are expressed in terms of boundary parameters $\{ \alpha ^{\mp}_l|l=1,2,3\}$ as following
\begin{eqnarray}
  c^{\mp}_x &=& e^{-i\pi(\sum_{l=1} \alpha^{\mp}_l -\frac{\tau}{2} )} \prod^3_{l=1} \frac{\sigma(\alpha^{\mp}_l -\frac{\tau}{2})}{\sigma(\alpha^{\mp}_l)} ,  \nonumber\\
  c^{\mp}_y &=& e^{-i\pi(\sum_{l=1} \alpha^{\mp}_l -\frac{1+\tau}{2} )} \prod^3_{l=1} \frac{\sigma(\alpha^{\mp}_l -\frac{1+\tau}{2})}{\sigma(\alpha^{\mp}_l)} ,  \nonumber \\
  c^{\mp}_z &=& \prod^3_{l=1} \frac{\sigma(\alpha^{\mp}_l -\frac{1}{2})}{\sigma(\alpha^{\mp}_l)} .
\end{eqnarray}
The QYBE (\ref{YBE}), RE (\ref{RE}) and dual RE (\ref{dual RE}) lead to the fact that the transfer matrices (\ref{tra}) with different spectral parameters commute with each other, i.e., $[t(u),t(v)]=0$.
Thus $t(u)$ serves as the generating function of the conserved quantities, which ensures the integrability of the system.
The model Hamiltonian (\ref{Hamil}) can be expressed in terms of the transfer matrix as
\begin{eqnarray}
  H = \frac{\sigma(\eta)}{\sigma'(0)} \left\{ \frac{\partial}{\partial u}\ln t(u)|_{u=0,\{ \theta_j=0 \}} -(N-1)\zeta(\eta) -2\zeta(2\eta)  \right\}, \label{Hamil2}
\end{eqnarray}
where $\sigma'(0)=\frac{\partial}{\partial u} \sigma(u)|_{u=0}$ and the elliptic function $\zeta(u)$ is defined by (\ref{zetaf}).

\subsection{The exact solutions}

Suppose the eigenvalue of the transfer matrix $t(u)$ is $\Lambda (u)$. By using the fusion technique, we have the following fusion relations \cite{cao2013off-xyz}
\begin{equation}\label{identity}
  \Lambda (\theta_j) \Lambda(\theta_j-\eta) = - \frac{\Delta_q(\theta_j ) \sigma(\eta) \sigma(\eta) }{\sigma(2\theta_j+\eta) \sigma(2\theta_j-\eta)}, \quad j=1,\cdots, N,
\end{equation}
where $\Delta_q(u)$ is the quantum determinant
\begin{eqnarray*}
  \Delta_q(u) &=& -\frac{\sigma(2u + 2\eta) \sigma(2u - 2\eta)}{\sigma(\eta) \sigma(\eta)} \prod_{\gamma=\pm} \prod^3_{l=1} \frac{\sigma(u+\alpha^{\gamma}_l)}{\sigma(\alpha^{\gamma}_l)} \frac{\sigma (u-\alpha^{\gamma}_l)}{\sigma(\alpha^{\gamma}_l)} \nonumber \\
  && \times \prod^N_{k=1} \frac{\sigma(u+\theta_k+\eta)\sigma(u+\theta_k-\eta) \sigma(u-\theta_k+\eta) \sigma(u-\theta_k-\eta)}{\sigma(\eta)\sigma(\eta) \sigma(\eta)\sigma(\eta)}.
\end{eqnarray*}
From the direct calculation, we also have the values of $\Lambda (u)$ at certain points
\begin{eqnarray}
  && \Lambda(0)= \frac{\sigma(2\eta)}{\sigma(\eta)} \prod^N_{k=1}\frac{\sigma(\eta+\theta_k)\sigma(\eta-\theta_k)}{\sigma(\eta)\sigma(\eta)}, \label{Lam0} \\
  && \Lambda(\frac{1}{2})= (-1)^{N+1} c^-_z c^+_z  \frac{\sigma(2\eta)}{\sigma(\eta)}  \prod^N_{k=1}\frac{\sigma(\eta+\frac{1}{2}+\theta_k)
  \sigma(\eta-\frac{1}{2}-\theta_k)}{\sigma(\eta)\sigma(\eta)}, \label{Lam1}\\
  && \Lambda(\frac{\tau}{2})= (-1)^{N+1} c^-_x c^+_x  \frac{\sigma(2\eta)}{\sigma(\eta)}    e^{-i\pi[ (N+3)\eta +\tau - 2\sum^N_{j=1} \theta_j] } \nonumber\\
  && \qquad \quad \times \prod^N_{k=1}\frac{\sigma(\eta+\frac{\tau}{2}+\theta_k)
  \sigma(\eta-\frac{\tau}{2}-\theta_k)}{\sigma(\eta)\sigma(\eta)}, \label{Lam2} \\
   && \Lambda(\frac{1+\tau}{2}) = (-1)^N  c^-_y c^+_y  \frac{\sigma(2\eta)}{\sigma(\eta)}   e^{-i\pi[(N+3)\eta +\tau -2\sum^N_{j=1} \theta_j] } \nonumber\\
  && \qquad \quad \times \prod^N_{k=1}\frac{\sigma(\eta+\frac{1+\tau}{2}+\theta_k)
  \sigma(\eta-\frac{1+\tau}{2}-\theta_k)}{\sigma(\eta)\sigma(\eta)}. \label{Lam3}
\end{eqnarray}
Besides, the eigenvalue $\Lambda (u)$ satisfies the following crossing symmetry
\begin{eqnarray}
\Lambda(-u-\eta)= \Lambda(u).  \label{cror}
\end{eqnarray}
The quasi-periodic properties of of $\Lambda (u)$ are
\begin{equation}\label{quasi-p}
 \Lambda(u+1)=\Lambda(u), \quad \Lambda(u+\tau)=e^{-2i\pi(N+3)(2u+\eta+\tau)}\Lambda(u).
\end{equation}

From the definitions of $R$-matrix (\ref{Rm}) and $K$-matrices (\ref{Kmm})-(\ref{Kpm}), we know that
\begin{equation}\label{analy}
 \textrm{$\Lambda(u)$, as an entire function of $u$, is an elliptic polynomial of degree $2N+6$.}
\end{equation}
Thus the constraints (\ref{identity})-(\ref{cror}) can completely determine the value of $\Lambda(u)$.
Considering the crossing symmetry (\ref{cror}) and quasi-periodic properties (\ref{quasi-p}),
we parameterize the $\Lambda(u)$ as
\begin{equation}\label{eiv}
  \Lambda(u)=\Lambda_0 \prod^{N+3}_{l=1} \sigma(u+z_l+\frac{\eta}{2}) \sigma(u - z_l+\frac{\eta}{2}),
\end{equation}
where $\{z_l |l=1,\cdots ,N+3  \}$ are the zero roots of the elliptic polynomial and $\Lambda_0$ is a coefficient.
The values of $N+4$ unknowns $\{z_l |l=1,\cdots,N+3  \}$ and $\Lambda_0$ are determined by the constraints (\ref{identity})-(\ref{Lam3})
with a given set of inhomogeneity parameters $\{\theta_j |j=1,\cdots,N\}$.

From (\ref{Hamil2}) and (\ref{eiv}), the eigen-energies of the Hamiltonian (\ref{Hamil}) are
\begin{eqnarray}
  E &=& \frac{\sigma(\eta)}{\sigma'(0)} \left\{ \frac{\partial}{\partial u}\ln \Lambda(u)|_{u=0} -(N-1) \frac{\sigma'(\eta)}{\sigma(\eta)} -2\frac{\sigma'(2\eta)}{\sigma(2\eta)}  \right\} \nonumber\\
    &=& \frac{\sigma(\eta)}{\sigma'(0)} \left\{  \sum^{N+3}_{l=1} \left( \frac{ \sigma'(z_l+\frac{\eta}{2})}{ \sigma(z_l+\frac{\eta}{2}) } - \frac{ \sigma'(z_l - \frac{\eta}{2})}{ \sigma( z_l - \frac{\eta}{2}) }  \right) - (N-1) \frac{\sigma'(\eta)}{\sigma(\eta)} - 2\frac{\sigma'(2\eta)}{\sigma(2\eta)}   \right\} . \label{Evalue}
\end{eqnarray}

\subsection{The Hermiticity}

To study the physical properties of the model, we should first consider the Hermiticity of the Hamiltonian (\ref{Hamil}), where the coupling constants $\{J_x, J_y, J_z\}$ and boundary fields $\{ h^{\mp}_x, h^{\mp}_y, h^{\mp}_z \}$ should be real.
In this paper, we take the modulus parameter $\tau$ to be pure imaginary and $\textrm{Im} (\tau) >0$. As a result, the constraint of real coupling constants gives that $\eta$ is real or is pure imaginary.

For convenience, we parameterize the boundary parameters $\{\alpha^{\mp}_j\}$ in the following forms
\begin{equation}\label{repara}
  \alpha^{\mp}_1 = \beta^{\mp}_1, \quad \alpha^{\mp}_2 = \beta^{\mp}_2 +\frac{\tau}{2}, \quad \alpha^{\mp}_3 = \beta^{\mp}_3  +\frac{1}{2}.
\end{equation}
To ensure the boundary fields $\{ h^{\mp}_x, h^{\mp}_y, h^{\mp}_z \}$ are real, the values of boundary parameters $\{ \beta^{\mp}_l|l=1,2,3\}$ should also satisfy some constraints.
Clearly, the boundary fields $\{h^{\mp}_x, h^{\mp}_y, h^{\mp}_z\}$ are invariant with the parameter changes $ \beta^{\mp}_j\rightarrow \beta^{\mp}_j+2 $ and $ \beta^{\mp}_j\rightarrow \beta^{\mp}_j+2\tau $, due to the periodicity of the elliptic function.
Without losing generality, we restrict the boundary parameters to the intervals $\Re(\beta^{\mp}_j) \in [0,2]$ and $\Im(\beta^{\mp}_j) \in [0,2 \frac{\tau}{i}]$.
Substituting Eq.(\ref{repara}) into (\ref{hfx})-(\ref{hfz}) and conducting the theoretical analysis of the elliptic functions, we determine the Hermitian regions of the boundary parameters.
These regions depend on the values of crossing parameter $\eta$. When $\eta$ is real, the Hermitian regions of the boundary parameters are
\begin{eqnarray}
 && \Re(\beta^{\mp}_1)\in(0,2), \qquad \Im(\beta^{\mp}_1)=0 \quad {\rm or} \quad  \Im(\beta^{\mp}_1)= \frac{\tau}{i},   \nonumber\\
 && \Re(\beta^{\mp}_2)\in[0,2],  \qquad  \Im(\beta^{\mp}_2)=0 \quad {\rm or} \quad  \Im(\beta^{\mp}_2)= \frac{\tau}{i},  \nonumber\\
 && \Re(\beta^{\mp}_3)=0 \quad {\rm or} \quad  \Re(\beta^{\mp}_3)=1,  \qquad  \Im(\beta^{\mp}_3)\in [0,2\frac{\tau}{i}].
\end{eqnarray}
When $\eta$ is pure imaginary, the Hermitian regions are
\begin{eqnarray}
  && \Re(\beta^{\mp}_1)=0  \quad {\rm or} \quad  \Re(\beta^{\mp}_1)=1,  \qquad \Im(\beta^{\mp}_1)\in(0,2\frac{\tau}{i}), \nonumber\\
  && \Re(\beta^{\mp}_2)\in[0,2], \qquad  \Im(\beta^{\mp}_2)=0 \quad {\rm or} \quad  \Im(\beta^{\mp}_2)= \frac{\tau}{i},  \nonumber\\
  && \Re(\beta^{\mp}_3)=0  \quad {\rm or} \quad  \Re(\beta^{\mp}_3)=1,  \qquad \Im(\beta^{\mp}_3)\in [0,2\frac{\tau}{i}].
\end{eqnarray}

Substituting Eqs.(\ref{eiv}) and (\ref{repara}) into (\ref{identity})-(\ref{Lam3}), one can easily find that the zero roots $ \{ z_l | l=1,\cdots, N+3 \}$ are invariant under the following parameter transformations
\begin{eqnarray}
  && \eta \rightarrow -\eta , \label{pt1} \\
  && \beta^{-}_j \rightleftharpoons \beta^{+}_j , \label{pt2}\\
  && \beta^{+}_j  \rightarrow  - \beta^{+}_j ~\textrm{and }~ \beta^{-}_j  \rightarrow  - \beta^{-}_j, \label{pt3}\\
  && \textrm{two of six boundary parameters change with } \beta^{\gamma}_j  \rightarrow  \beta^{\gamma}_j+1, \label{pt41}\\
  && \textrm{two of six boundary parameters change with } \beta^{\gamma}_j  \rightarrow  \beta^{\gamma}_j +\tau,\label{pt42}
\end{eqnarray}
where $\{j=1,2,3\}$ and $\{\gamma=\pm\}$. Then, the eigen-energies are invariant under the transformations (\ref{pt1})-(\ref{pt42}).
Further, from the Eq.(\ref{pt2}), we can only consider the case of $|\beta^{+}_j| \geq |\beta^{-}_j|$.

The transformations (\ref{pt2})-(\ref{pt42}) allow us to restrict the boundary parameters to the following compact regions as well as some parameter changes are necessary to ensure that all possible parameters are taken in account.
If $\eta$ is real, we first choose the region as
\begin{eqnarray}\label{r-in}
  && \Re(\beta^{\mp}_1) \in (0,\frac{1}{2}], \quad \Im(\beta^{\mp}_1)=0, \nonumber\\
  && \Re(\beta^{\mp}_2) \in [0,\frac{1}{2}], \quad \Im(\beta^{\mp}_2) = 0, \nonumber\\
  && \Re(\beta^{\mp}_3) = 0, \quad \Im(\beta^{\mp}_3) \in [0,\frac{\tau}{2i}].
\end{eqnarray}
Then we take following transformations
\begin{equation}\label{r-trans}
(\textrm{i}) \; \;  \beta^{+}_1  \rightarrow  1- \beta^{+}_1, \quad (\textrm{ii})  \; \; \beta^{+}_1  \rightarrow \beta^{+}_1 +1, \quad (\textrm{iii}) \; \;  \beta^{+}_1  \rightarrow \beta^{+}_1 +\tau,
\end{equation}
and all the regions of boundary parameters are covered.

If $\eta$ is pure imaginary, we can choose the region as
\begin{eqnarray}\label{i-in}
  && \Re(\beta^{\mp}_1) =0, \quad \Im(\beta^{\mp}_1)\in (0,\frac{\tau}{2i}], \nonumber\\
  && \Re(\beta^{\mp}_2) \in [0,\frac{1}{2}], \quad \Im(\beta^{\mp}_2) = 0, \nonumber\\
  && \Re(\beta^{\mp}_3) = 0, \quad \Im(\beta^{\mp}_3) \in [0,\frac{\tau}{2i}].
\end{eqnarray}
Then taking the following transformations
  \begin{equation}\label{i-trans}
  (\textrm{i}) \beta^{+}_1  \rightarrow  \tau- \beta^{+}_1, \quad (\textrm{ii})\beta^{+}_1  \rightarrow \beta^{+}_1 +1, \quad (\textrm{iii}) \beta^{+}_1  \rightarrow \beta^{+}_1 +\tau,
  \end{equation}
we can cover the rest regions of boundary parameters. We should note that after taking the transformations (\ref{r-trans}) and (\ref{i-trans}), the boundary fields change as
\begin{equation}\label{ftrans}
  ( h^{+}_x, h^{+}_y, h^{+}_z ) \Rightarrow \left\{
                                                      \begin{array}{cc}
                                                        (h^{+}_x, h^{+}_y,- h^{+}_z ), & \textrm{for}\quad \beta^{+}_1 \rightarrow 1 -\beta^{+}_1,\\
                                                        (- h^{+}_x, h^{+}_y, h^{+}_z ), & \textrm{for}\quad \beta^{+}_1 \rightarrow \tau -\beta^{+}_1,\\
                                                        (- h^{+}_x, -h^{+}_y, h^{+}_z ), & \textrm{for}\quad \beta^{+}_1 \rightarrow \beta^{+}_1+1 , \\
                                                        (h^{+}_x,- h^{+}_y,- h^{+}_z ), & \textrm{for}\quad  \beta^{+}_1 \rightarrow \beta^{+}_1+\tau,
                                                      \end{array}
                                                    \right.
\end{equation}
which means that the transformation (\ref{r-trans}) and (\ref{i-trans}) are equivalent to the flipping of certain components of the boundary magnetic fields.

Next, we study the thermodynamic limit of the system for both real and imaginary values of the parameter $\eta$.
If $\eta$ is real, the Hamiltonian (\ref{Hamil}) exhibits the periodicity $H(\eta+2)=H(\eta)$. Meanwhile, from the property (\ref{pt1}),
we conclude that only the interval $\eta \in (0,1)$ are necessary.
While if $\eta$ is pure imaginary, from the quasi-periodicity of the Hamiltonian $H(\eta+2\tau)=e^{-4i\pi(\eta+\tau)}H(\eta)$, we need only consider the
interval $\eta \in (0,\tau)$.

We should note that the degenerate points $\eta=0,1, \tau$ are excluded in this paper, because the model reduces to the isotropic XXX spin chain with free open boundaries at these points.

\section{Thermodynamic limit for the real $\eta$}
\label{sec-r}
\setcounter{equation}{0}

\subsection{ Patterns of zero roots}

For the real $\eta$, we examine the distribution patterns of zero roots $\{ z_l \}  \equiv \{ i \bar{z}_l \}$.
From the intrinsic properties of $R$-matrix and reflection matrices, we find that $t(u)^{\dag} =t(u^{\ast})$ and $\Lambda(u) = \Lambda^{*}(u^{*})$ if
the inhomogeneity parameters are pure imaginary $ \{\theta_j \} \equiv i \{\bar{\theta}_j \}$.
Then we deduce that if $\bar{z}_l$ is a root of $\Lambda(u)$, there must be another root $\bar{z}_j$ satisfy
\begin{equation}\label{period-r}
 \bar{z}_j=\bar{z}^{\ast}_l + m_1 \frac{\tau}{i} +m_2 i, \quad m_1, m_2 \in \mathbb{Z}.
\end{equation}
According to Eq.(\ref{period-r}), we conclude that the roots $\{\bar{z}_l\}$ can be classified into:
(i) real $\pm \bar{z}_l$; (ii) on the lines $\pm \frac{i}{2}$ (please note that its conjugate shifted by $i$ becomes itself); (iii) $(\pm \bar{z}_l, \pm \bar{z}^{\ast}_l)$ conjugate pairs and (iv) boundary strings induced by the boundary fields.
Eq.(\ref{period-r}) also allows us to fix the roots in the interval $\Im(\bar{z}_l)\in [-\frac{1}{2},\frac{1}{2}]$ and $\Re(\bar{z}_l)\in [-\frac{\tau}{2i},\frac{\tau}{2i}]$.

Now, we should determine the detailed form of the conjugate pairs in the cases (iii) and (iv), which can be achieved by
employing the method outlined in \cite{le2021root}.
If the crossing parameter $\eta$ takes the following discrete values \cite{wang2015off,yang2006t,zhang2022invariant}
\begin{eqnarray}
 && \eta_{L,K} =   \frac{2L}{2M-N+1}\tau + \frac{2K}{2M-N+1} -\sum_{\gamma=\pm} \sum_{j=1}^{3} \frac{ \varepsilon^{\gamma}_j \alpha^{\gamma}_j }{2M-N+1} ,  \nonumber \\
 && \prod\limits_{\gamma=\pm} \prod\limits_{j=1}^{3}  \varepsilon^{\gamma}_j =-1, \quad  \varepsilon^{\gamma}_j =\pm 1, \quad L,K\in \mathbb{Z}, \label{eta-discrete}
\end{eqnarray}
the eigenvalue $\Lambda(u)$ can also be expressed by the homogeneous $T-Q$ relation
\begin{equation}\label{T-Q}
  \Lambda (u) = a(u) \frac{Q(u-\eta)}{Q(u)} + d(u) \frac{Q(u+\eta)}{Q(u)}.
\end{equation}
Here $Q(u)$ is an elliptic polynomial
\begin{equation}
  Q(u)= \prod^{M}_{l=1} \frac{\sigma(u-u_l)\sigma(u+u_l+\eta)}{\sigma(\eta) \sigma(\eta)},
\end{equation}
and $\{u_l\}$ are the Bethe roots. The function $a(u)$ and $d(u)$ are
\begin{eqnarray}
  a(u) &=& - e^{-4i\pi L u} \frac{\sigma(2u+2\eta)}{\sigma(2u+\eta)} \prod_{\gamma=\pm} \prod_{l=1}^{3} \frac{\sigma(u- \varepsilon_l^{\gamma}\alpha_l^{\gamma} )}{\varepsilon_j^{\gamma}\alpha_j^{\gamma} } \nonumber\\
  && \times \prod^{N}_{j=1} \frac{\sigma(u+\theta_j+\eta) \sigma(u-\theta_j +\eta)}{\sigma(\eta) \sigma(\eta)},  \\
  d(u) &=& a(-u-\eta) . \label{d-function}
\end{eqnarray}
In the thermodynamic limit $N\rightarrow \infty$, the third term of $\eta_{L,K}$ (\ref{eta-discrete}) becomes zero.
By adjusting the integers  $L$ and $K$ appropriately,  $\eta_{L,K}$ can attain any complex number. Moreover, for real $\eta$, we can set $L=0$ to obtain the desired result.
Putting $u_l \equiv \lambda_l i -\frac{\eta}{2}$ and considering the homogeneous limit $\{\theta_j\rightarrow 0\}$, the Bethe roots $\{\lambda_l\}$ satisfy the Bethe ansatz equations (BAEs)
\begin{eqnarray}
   && - \frac{\sigma(i(2\lambda_j-\eta i))}{\sigma(i(2\lambda_j+\eta i))} \prod_{\gamma=\pm}\prod_{k=1}^3 \frac{\sigma(i(\lambda_j+\varepsilon_k^{\gamma} \alpha_{k}^{\gamma}i +\frac{\eta}{2}i))}{\sigma(i(\lambda_j - \varepsilon_k^{\gamma} \alpha_{k}^{\gamma}i - \frac{\eta}{2}i))} \left(  \frac{\sigma(i(\lambda_j-\frac{\eta}{2}i))}{\sigma(i(\lambda_j+\frac{\eta}{2}i))} \right)^{2N} \nonumber \\
   =&& \prod_{l=1}^N \frac{\sigma(i(\lambda_j- \lambda_l -\eta i))\sigma(i(\lambda_j+ \lambda_l -\eta i)) }{\sigma(i(\lambda_j- \lambda_l +\eta i))\sigma(i(\lambda_j+ \lambda_l +\eta i))}, \quad j=1,\cdots,N. \label{rBAEs}
\end{eqnarray}
For a complex Bethe root $\lambda_j$ with an imaginary part, we readily have $\left|  \frac{\sigma(i(\lambda_j-\frac{\eta}{2}i))}{\sigma(i(\lambda_j+\frac{\eta}{2}i))} \right|\neq 1 $.
This indicates that in the thermodynamic limit $N\rightarrow \infty$, the left hand side of  Eq.(\ref{rBAEs}) tends to infinity or zero exponentially.
To keep Eq.(\ref{rBAEs}) holding, the right hand side must tend to infinity or zero with the same order.
Then we arrive at that the Bethe roots satisfy the string hypothesis \cite{takahashi1999thermodynamics, xin2020thermodynamic, Takahashi1972one}
\begin{equation}\label{string-r}
  \lambda_{j,k} = x_{j}+(\frac{n_j+1}{2} -k)\eta i +\frac{1-\nu_j}{4}i +O(e^{-\delta N}), \quad 1\leq k \leq n_j,
\end{equation}
where $x_j$ is the position of the $j$-string on the real axis, $k$ means the $k$th Bethe roots in $j$-string, $O(e^{-\delta N})$ means the finite size correction,  $n_j$ is the length of $j$-string, and $\nu_j=\pm 1$ denotes the parity of $j$-string.
The center of $j$-string is the real axis if $\nu_j=1$, while the center of $j$-string is the line with fixed imaginary party $\frac{i}{2}$ in the complex plane if $\nu_j=-1$.

The structure of zero roots of eigenvalue $\Lambda(u)$ can be obtained by Eq.(\ref{T-Q}).
Substituting $\{u= i\bar{z}_j -\frac{\eta}{2} \}$ into (\ref{T-Q}), we obtain the relation between zero roots $\{\bar{z}_j\}$ and Bethe roots $\{\lambda_j\}$
\begin{eqnarray}
   && - \frac{\sigma(i(2 \bar{z}_j -\eta i))}{\sigma(i(2 \bar{z}_j +\eta i))} \prod_{\gamma=\pm}\prod_{k=1}^3 \frac{\sigma(i(\bar{z}_j +\varepsilon_k^{\gamma} \alpha_{k}^{\gamma}i +\frac{\eta}{2}i))}{\sigma(i(\bar{z}_j - \varepsilon_k^{\gamma} \alpha_{k}^{\gamma}i - \frac{\eta}{2}i))} \left(  \frac{\sigma(i(\bar{z}_j-\frac{\eta}{2}i))}{\sigma(i(\bar{z}_j +\frac{\eta}{2}i))} \right)^{2N} \nonumber \\
   =&& \prod_{l=1}^N \frac{\sigma(i(\bar{z}_j- \lambda_l -\eta i))\sigma(i(\bar{z}_j+ \lambda_l -\eta i)) }{\sigma(i(\bar{z}_j - \lambda_l +\eta i))\sigma(i(\bar{z}_j + \lambda_l +\eta i))}, \quad j=1,\cdots,N+3.\label{rBAE-zero}
\end{eqnarray}
We should note the roots of functions $\Lambda(u)$ and $Q(u)$ could not be equal, i.e. $\bar{z}_j \neq \lambda_j$. Thus from the structure of Bethe roots $\{ \lambda_{j,k}\}$ (\ref{string-r}), we obtain the patterns of zero roots $\{\bar{z}_j\}$ as
\begin{equation}
  \bar{z}_{j} = x_{j} \pm \frac{n_j+1}{2}\eta i +\frac{1-\nu_j}{4}i +O(e^{-\delta N}).
\end{equation}
For clarity, we present two simple patterns of zero roots as below: for $n_j=1, \nu_j=1$, the root pattern is $\bar{z}_{j}=x_{j} \pm \eta i $, and for $n_j=1, \nu_j=-1$, the root pattern is $\bar{z}_{j}=x_{j} \pm (\frac{1}{2} - \eta)i $ which is adjusted to ensure that $\Im(\bar{z}_{j})\in [-\frac{1}{2},\frac{1}{2}]$ by shifting $i$.

From the singularity analysis of left hand side of Eq.(\ref{rBAE-zero}), we obtain that the boundary strings $\{ w_l^{\gamma} \}$ take the forms of
\begin{equation}\label{discrete-r}
w_l^{\gamma} = \pm \left( \frac{\eta}{2} + \varepsilon_l^{\gamma} \alpha_{l}^{\gamma} \right) i, \quad \gamma =\pm.
\end{equation}

By using the numerical exact diagonalization, we obtain the zero roots of the eigenvalue $\Lambda(u)$ with $\eta\in(0,1)$.
We find that the distribution of roots in the interval $\eta \in (0,\frac{1}{2})$ and that in the interval $\eta \in (\frac{1}{2},1)$ are different.
Thus we should discuss them separately.

\subsection{Surface energy and excitation with $\eta \in (\frac{1}{2}, 1)$}
\label{rb}

\subsubsection{Results in the boundary parameters region (\ref{r-in})}

\begin{figure}[htbp]
\centering
\subfigure[$N=8$]{
\includegraphics[width=4.5cm]{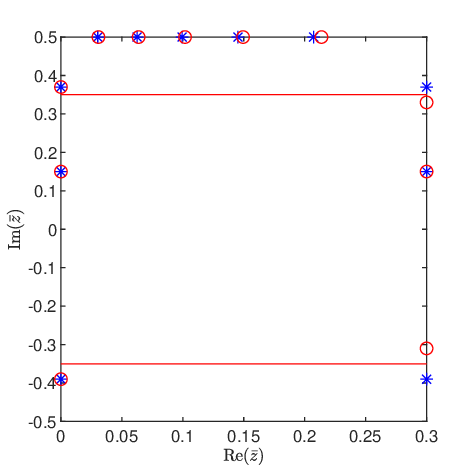}
}
\subfigure[$N=9$]{
\includegraphics[width=4.5cm]{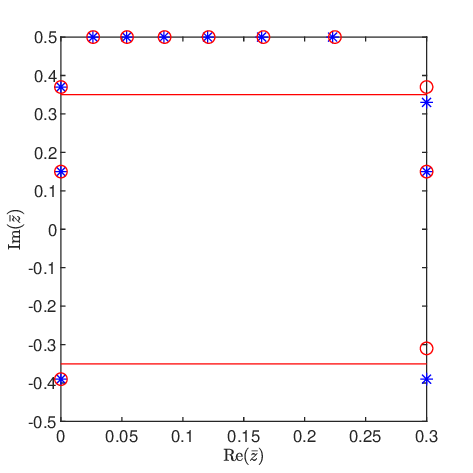}
}
\caption{Exact numerical results of the zero roots at the ground and first excited states for (a) $N=8$ and (b) $N=9$. Here $\tau=0.6i$, $\eta=0.7$, $\beta^{+}_1=0.04, \beta^{-}_1=0.02, \beta^{+}_2=0.04, \beta^{-}_2=0.02, \beta^{+}_3=0.04i$ and $\beta^{-}_3=0.03i$.
The asterisks indicate the roots at the ground state and the circles indicate the results at the first excited state.
For the clearness, we also draw the lines with fixed imaginary parts $\pm \frac{\eta}{2} i$.
}\label{rb-gedis}
\end{figure}

We first consider the region (\ref{r-in}) of boundary parameters. The numerical results of the zero roots of function $\Lambda(u)$ with finite system size at the ground state and at the first excited state are shown in Fig.\ref{rb-gedis}.
It can be seen that the zero roots in the bulk are located on the line $\frac{i}{2}$ with the form of $ \{ \bar{z}_l= x_l+\frac{i}{2} | l=1,\cdots,n_1 \}$ where $x_l$ is real and $x_l\in [0, \frac{\tau}{2i}]$.
These roots are continuously distributed in the thermodynamic limit.
The $n_2=N+3-n_1$ discrete boundary strings $ \{w_t| t=1,\cdots,n_2 \}$ are the complex roots in the imaginary axis and the vertical line with $\Re(\bar{z})=\frac{\tau}{2i}$.

Based on the zero roots patterns derive above, the physical quantities such as the ground state energy and excitation energy in the thermodynamic limit can be computed.
The key point of the process is to introduce a proper set of inhomogeneity parameters, which serve as the auxiliary tool for the computation and will be taken to zero finally.
For the real $\eta$, we choose all the inhomogeneity parameters to be imaginary, i.e., $\{\theta_j\}=\{\bar{\theta}_j i\}$.
Such a choice does not change the patterns of the roots but the root density.

In the thermodynamic limit, we assume that the inhomogeneity has a density $\varrho(\bar{\theta})\sim 1/N(\bar{\theta}_{j}-\bar{\theta}_{j-1})$.
Taking the logarithm of Eq.(\ref{identity}) and making the difference of the equations for $\bar{\theta}_{j}$ and $\bar{\theta}_{j-1}$, by omitting the $O(N^{-1})$ terms we readily have
\begin{eqnarray}
&&  N \int^{\frac{\tau}{2i}}_{-\frac{\tau}{2i}} A_{ \frac{1-\eta}{2}i}( u -x) \rho(x) dx    + \sum^{n_2}_{t=1} \left(  A_{w_t-\frac{\eta}{2}i} (u ) + A_{w_t+\frac{\eta}{2}i}  (u ) \right)  \nonumber\\
&=&  A_{\eta i}(u) +A_{\eta i + \frac{i}{2}}(u ) +A_{\eta i +\frac{\tau}{2}i}(u ) +A_{\eta i +\frac{\tau-1}{2}i}(u )   \nonumber\\
   && - A_{\frac{\eta}{2}i}(u) - A_{\frac{\eta+1}{2}i}(u ) -A_{\frac{\eta+\tau}{2}i}(u ) -A_{\frac{\eta+1+\tau }{2}i}(u )   \nonumber\\
   && + N \int^{\frac{\tau}{2i}}_{-\frac{\tau}{2i}} \left\{ A_{\eta i}( u +\bar{\theta} ) +A_{\eta i}( u -\bar{\theta} ) \right\} \varrho (\bar{\theta}) d \bar{\theta}  + \sum_{\gamma=\pm} \sum^3_{l=1} A_{ \alpha^{\gamma}_l i}( u ), \label{BAr}
\end{eqnarray}
where $\rho(x)$ is the density of the bulk zero roots located on the line $\frac{i}{2}$,
$A_{\gamma}(u)$ is a periodic function with the form
\begin{eqnarray}
A_{\gamma}(u) = \frac{\sigma'(i(u- {\gamma}))}{\sigma(i(u- {\gamma}))} +\frac{\sigma'(i(u+ {\gamma}))}{\sigma(i(u+ {\gamma}))} - \frac{4\pi}{\tau}u ,  \label{Afr}
\end{eqnarray}
and ${\gamma}$ is the complex number with $\Im({\gamma})\in [-1,1]$ and $\Re ({\gamma})\in [-\frac{\tau}{2i}, \frac{\tau}{2i}]$.

Taking homogeneous limit $\varrho(\bar{\theta})\rightarrow \delta (\bar{\theta})$, the density $\rho(x)$ of the corresponding homogeneous system can be obtained via the following Fourier transform
\begin{eqnarray}
 \tilde{F}(k)= \int^{T}_{-T} F(x)e^{-ik\frac{\pi}{T}x} dx , \quad
  F(x) = \frac{1}{2T} \sum^{\infty}_{k=-\infty} \tilde{F}(k)e^{ik\frac{\pi}{T}x}, \quad k \in \mathbb{Z},\label{F}
\end{eqnarray}
where $F(x)$ is a periodic function and $2T$ is the periodicity.
The solution of the density is
\begin{eqnarray}
 N   \tilde{\rho}(k)   &=& \frac{1}{\tilde{A}_{ \frac{1-\eta}{2}i}( k)}  \left\{ 2N \tilde{A}_{\eta i}( k )  + \sum_{\gamma=\pm} \sum^3_{l=1} \tilde{A}_{ \alpha^{\gamma}_l i}( k ) + \tilde{A}_{\eta i}(k)  +\tilde{A}_{\eta i -\frac{i}{2}}(k )\right. \nonumber\\
   &&  +\tilde{A}_{\eta i +\frac{\tau}{2}i}(k ) +\tilde{A}_{\eta i +\frac{\tau-1}{2}i}(k )  - \tilde{A}_{\frac{\eta}{2}i}(k) - \tilde{A}_{\frac{\eta+1}{2}i}(k ) -\tilde{A}_{\frac{\eta+\tau}{2}i}(k )  \nonumber\\
   && \left. -\tilde{A}_{\frac{\eta+1+\tau }{2}i}(k ) -  \sum^{n_2}_{t=1} \left(  \tilde{A}_{w_t-\frac{\eta}{2}i} (k ) + \tilde{A}_{w_t+\frac{\eta}{2}i}  (k ) \right)  \right\}, \label{densityF2}
\end{eqnarray}
where $\tilde{A}_{\gamma}(k)$ is the Fourier transformation of function $A_{\gamma}(u)$
\begin{equation}\label{AfrF}
  \tilde{A}_{\gamma}(k) = \left\{ \begin{array}{cc}
                      - 2\pi   \left(  \cosh(\frac{ik\pi}{\tau}  2 {\gamma} i ) \coth(\frac{ik\pi}{\tau}) + \sI(\gamma) \sinh(\frac{ik\pi}{\tau} 2 \gamma i )  \right) , & k\neq 0, \\
                     0, & k=0,
                   \end{array}\right.
\end{equation}
and
\begin{equation}\label{sI}
  \sI(\gamma)= \left\{ \begin{array}{cc}
                                  1, & \Im(\gamma)>0, \\
                                  0, & \Im(\gamma)=0, \\
                                  -1, & \Im(\gamma)<0.
                                \end{array} \right.
\end{equation}

Substituting the patterns of zero roots into Eq.(\ref{Evalue}) and taking the thermodynamic limit, we obtain the eigen-energy $E$ as
\begin{eqnarray}
E &=&  -\frac{N}{2} \frac{\sigma(\eta)}{\sigma'(0)} \int^{\frac{\tau}{2i}}_{-\frac{\tau}{2i}}B_{ \frac{1-\eta}{2}i}  (x) \rho(x) dx - N  \frac{\sigma'(\eta)}{\sigma'(0)} + \frac{\sigma'(\eta)}{\sigma'(0)} - 2  \frac{\sigma(\eta)}{\sigma'(0)} \frac{\sigma'(2\eta)}{\sigma(2\eta)} \nonumber \\
&& + \frac{\sigma(\eta)}{\sigma'(0)} \frac{1}{2} \sum^{n_2}_{t=1}  \left(  B_{w_t+\frac{\eta}{2}i }(0)- B_{w_t-\frac{\eta}{2}i }(0)  \right)   \nonumber \\
&=& - \frac{\sigma(\eta)}{\sigma'(0)} \frac{N}{2} \frac{i}{\tau}  \sum^{\infty}_{k=- \infty} \tilde{B}_{ \frac{1-\eta}{2}i} (k) \tilde{\rho}(k)  - N  \frac{\sigma'(\eta)}{\sigma'(0)} + \frac{\sigma'(\eta)}{\sigma'(0)} - 2  \frac{\sigma(\eta)}{\sigma'(0)} \frac{\sigma'(2\eta)}{\sigma(2\eta)}   \nonumber \\
&& + \sum^{n_2}_{t=1} \frac{\sigma(\eta)}{\sigma'(0)} \frac{1}{2} \frac{i}{\tau} \sum^{\infty}_{k=- \infty } \left( \tilde{B}_{w_t+\frac{\eta}{2}i }(k) - \tilde{B}_{w_t-\frac{\eta}{2}i }(k)   \right), \label{rbE}
\end{eqnarray}
where $B_{\gamma}(u)$ is a periodic function with the form
\begin{eqnarray}
  B_{\gamma}(u) = \frac{\sigma'(i(u- {\gamma}))}{\sigma(i(u-{\gamma}))} - \frac{\sigma'(i(u+ {\gamma}))}{\sigma(i(u+{\gamma}))}, \label{Bfr}
\end{eqnarray}
and $\tilde{B}_{\gamma}(k)$ is the Fourier transformations of $B_{\gamma}(u)$
\begin{equation}\label{BfrF}
  \tilde{B}_{\gamma}(k) = \left\{ \begin{array}{cc}
                      2\pi   \left(  \sinh(\frac{ik\pi}{\tau}  2 \gamma i ) \coth(\frac{ik\pi}{\tau}) + \sI(\gamma) \cosh(\frac{ik\pi}{\tau} 2 \gamma i )  \right) , & k\neq 0, \\
                      2\pi   \left(   2 \gamma i  + \sI(\gamma)  \right) , & k=0.
                    \end{array}\right.
\end{equation}

Substituting the density (\ref{densityF2}) into (\ref{rbE}), the energy $E$ is solved as
\begin{eqnarray}
E = e_{r} N + E^{f}_{r} + E^{+}_{r} +E^{-}_{r} + \sum^{n_2}_{t=1} E^{w}_{r}(w_t), \label{rE}
\end{eqnarray}
where $e_{r}$ is the energy density in the bulk
\begin{eqnarray}
 e_{r} = - \frac{\sigma(\eta)}{\sigma'(0)} \left\{  4\pi \frac{i}{\tau} \sum^{\infty}_{k= 1} \tanh( \frac{ik\pi}{\tau}\eta )  \frac{  \cosh( \frac{ik\pi}{\tau}( 1- 2\eta ))  }{\sinh(\frac{ik\pi}{\tau})}  + \frac{i}{\tau} 2 \pi \eta   + \frac{\sigma'(\eta)}{\sigma(\eta)}  \right\} , \label{rbEden}
\end{eqnarray}
$E^{f}_{r}$ indicate the contributions of free open boundaries
\begin{eqnarray}
  E^{f}_{r} &=& -2\pi \frac{\sigma(\eta)}{\sigma'(0)}  \frac{i}{\tau}  \sum^{\infty}_{k=1} \tanh(\frac{ik\pi}{\tau}\eta) \frac{1 + \cos(k\pi)}{\sinh(ik\pi/\tau)}  \left\{ \cosh(\frac{ik\pi}{\tau}(1-2\eta))   \right. \nonumber\\
 && \left. + \cosh(\frac{ik\pi}{\tau}(1-|1-2\eta|)) - \cosh(\frac{ik\pi}{\tau}(1-\eta ))  - \cosh(\frac{ik\pi}{\tau} \eta )    \right\} \nonumber\\
  && +  \frac{\sigma'(\eta)}{\sigma'(0)} - 2  \frac{\sigma(\eta)}{\sigma'(0)} \frac{\sigma'(2\eta)}{\sigma(2\eta)}, \label{rEfree}
\end{eqnarray}
and $E^{\pm}_{r}$, $E^r(w_t)$ are the energies induced by the boundary magnetic fields
\begin{eqnarray}
E^{\pm}_{r} &=& -  \frac{\sigma(\eta)}{\sigma'(0)} 2\pi \frac{i}{\tau} \sum^{\infty}_{k=1}   \frac{  \tanh(ik\pi\eta/\tau) }{\sinh(ik\pi/\tau)}   \left\{ \cosh( 2 \frac{ik\pi}{\tau} (\frac{1}{2}-\beta^{\pm}_1 )) \right. \nonumber\\
&& \left. + \cosh( 2 \frac{ik\pi}{\tau}(\frac{1}{2}-\beta^{\pm}_2 ))\cos(k\pi)   + \cosh( 2\frac{ik\pi}{\tau} \beta^{\pm}_3  ) \right\}   -  \frac{\sigma(\eta)}{\sigma'(0)} \frac{i}{\tau} 3\pi \eta, \label{rbEpara}\\
E^{w}_{r}(w_t)& =&  \left( \sI( w_t +\frac{\eta}{2}i ) - \sI( w_t -\frac{\eta}{2}i )   \right) \frac{\sigma(\eta)}{\sigma'(0)}  \frac{i\pi}{\tau} \sum^{\infty}_{k= -\infty }  \frac{ \cosh( \frac{ik\pi}{\tau} 2w_ti ) }{ \cosh( \frac{ik\pi}{\tau} \eta) } . \label{rbEroot}
\end{eqnarray}
From Eq.(\ref{rbEroot}), we see that only the discrete roots with $\Im(w_t)\in [-\frac{\eta}{2},\frac{\eta}{2}]$ can contribute the non-zero boundary energy to the system.

Next, we should study the discrete roots. From the numerical simulation as shown in Fig.\ref{rb-gedis}, we find that there exist two discrete zero roots
\begin{equation} \label{rb-law12}
 w_1=  \frac{1- \eta}{2} i , \quad  w_2= \frac{\tau}{2i} + \frac{1- \eta}{2} i,
\end{equation}
which do not depend on the boundary parameters and the ground or first excited states.
According to Eq.(\ref{rbEroot}), above two roots will contribute a non-zero eigen-energy.

The remaining discrete roots can be divided into two types. The first kind of discrete roots are the complex numbers located long the line ${\rm Re}(\bar z) =0 $ induced by the parameter $\beta^{\pm}_1$
and the second ones are that located along the line ${\rm Re}(\bar z) =\frac{\tau}{2i}$ induced by the parameters $\beta^{\pm}_2$.
Substituting Eq.(\ref{repara}) into (\ref{discrete-r}), we get the forms of first kind of discrete roots as $(\frac{\eta}{2} + \varepsilon^{-}_{1} \beta^{-}_1)i$ and $-(\frac{\eta}{2} + \varepsilon^{+}_{1} \beta^{+}_1)i$,
where the minimum distance between them is $2\eta-1$.
The second kind of discrete roots are $\frac{\tau}{2i} + (\frac{\eta}{2} + \varepsilon^{-}_{2} \beta^{-}_2)i$ and $\frac{\tau}{2i} -(\frac{\eta}{2} + \varepsilon^{+}_{2} \beta^{+}_2)i$,
where the minimum distance between them also is  $2\eta-1$.
With the changing of boundary parameters
$\beta^{\pm}_1$ and $\beta^{\pm}_2$, some discrete roots move and locate on the line $\frac{i}{2}$. Therefore,
the first kind of boundary strings take the forms
\begin{eqnarray}
  w_3 =  \min \left\{ \frac{\eta}{2} + \beta^{-}_1, \frac{1}{2} \right\} i,  \qquad
  w_4 = - \min \left\{ \frac{\eta}{2} + \beta^{+}_1, \frac{1}{2} \right\} i.  \label{rb-law4}
\end{eqnarray}
These boundary strings contribute nothing to the energy according to Eq.(\ref{rbEroot}).
It should be noted that as shown in Fig.\ref{rb-gedis}, these boundary strings keep unchanged at both the ground state and the first excited state, and are independent of the parities of the sites number $N$.

However, the second kind of boundary strings  depends on the states of the system and the parities of $N$.
At the ground state, the boundary strings are
\begin{eqnarray}
  w_{5g}= \frac{\tau}{2i} + \min \left\{ \frac{\eta}{2} + \beta^{-}_2, \frac{1}{2} \right\} i,  \qquad
  w_{6g}= \frac{\tau}{2i} - \min \left\{ \frac{\eta}{2} + \beta^{+}_2, \frac{1}{2} \right\} i, \label{rb-law6-eveng}
\end{eqnarray}
for even $N$ and are
\begin{eqnarray}
  w_{5g} = \frac{\tau}{2i} + \max \left\{ \frac{\eta}{2} - \beta^{-}_2, 0 \right\} i, \qquad
  w_{6g} = \frac{\tau}{2i} - \min \left\{ \frac{\eta}{2} + \beta^{+}_2, \frac{1}{2} \right\} i. \label{rb-law6-oddg}
\end{eqnarray}
for odd $N$.

Substituting the discrete roots (\ref{rb-law12}) and  boundary strings (\ref{rb-law6-eveng}) or (\ref{rb-law6-oddg}) into (\ref{rE}), we obtain the ground state energy as
\begin{equation}\label{Egr}
E^{g}_{r} = e_{r}N + E^{f}_{r} +E^{+}_{r} +E^{-}_{r} +  E^{w}_{r}(w_1) +  E^{w}_{r}(w_2) +  E^{w}_{r}(w_{5g}) +  E^{w}_{r}(w_{6g}).
\end{equation}
The definition of the surface energy is defined as $E^s_{r} \equiv E^{g}_{r} -E^{gp}_{r}$, where
$E^{gp}_{r}$ is the ground state energy of the periodic XYZ spin chain which reads $E^{gp}_{r}= Ne_{r} +\frac{1}{2}(1- (-1)^N) E^{w}_{r}(\frac{\tau}{2i})$.
Then the surface energy of present system is
\begin{eqnarray}
E^s_{r} &=& E^{f}_{r} +E^{+}_{r} +E^{-}_{r} +  E^{w}_{r}(w_1) +  E^{w}_{r}(w_2) +  E^{w}_{r}(w_{5g}) \nonumber\\
 &&  +  E^{w}_{r}(w_{6g})- \frac{1}{2}(1- (-1)^N ) E^{w}_{r}(\frac{\tau}{2i}). \label{rbEs}
\end{eqnarray}

Next, we consider the elementary excitation. By comparing the distribution of zero roots at the ground state and that at the first excited state,
we find that the boundary strings located on the vertical line $\Re(\bar{z})=\frac{\tau}{2i}$ are changed, while others are similar due to the
reconstruction of Fermi sea, which can seen clearly from Fig.\ref{rb-gedis}.
Furthermore, the structure of boundary strings depends on the parity of $N$.
If $N$ is even, the boundary strings are
\begin{eqnarray}
  w_{5e} = \frac{\tau}{2i} + \phi_1 i, \qquad  w_{6e} = \frac{\tau}{2i} - \max \left\{ \frac{\eta}{2}-\beta^{+}_2, 2\eta -1 - \phi_1 , 0 \right\} i, \label{rb-law6-evene}
\end{eqnarray}
with $\phi_1= \max \left\{ \frac{\eta}{2}-\beta^{-}_2, \eta-\frac{1}{2} \right\}$.
As illustrated in Fig.\ref{rb-gedis}(a), the boundary strings $w_{5g}$ and $w_{6g}$ (\ref{rb-law6-eveng}) at the ground state are excited to their symmetrical positions with respect to the red lines $\Im(\bar{z})=\frac{\eta}{2}$
and $\Im(\bar{z})=- \frac{\eta}{2}$, respectively.
The boundary strings $w_{5e}$ and $w_{6e}$ (\ref{rb-law6-evene}) are located in the region within two lines $\Im(\bar{z})=\pm\frac{\eta}{2}$, and contribute the non-zero eigen-energies.

If $N$ is odd, the boundary strings are
\begin{eqnarray}
  w_{5e} = \frac{\tau}{2i} + \min \left\{ \frac{\eta}{2} + \beta^{-}_2, \frac{1}{2} \right\} i, \quad
  w_{6e} = \frac{\tau}{2i} - \max \left\{ \frac{\eta}{2}-\beta^{+}_2, \frac{3\eta}{2}-1 -\beta^{-}_2, 0 \right\} i. \label{rb-law6-odde}
\end{eqnarray}
As illustrated in Fig.\ref{rb-gedis}(b), we see that the boundary strings $w_{5g}$ and $w_{6g}$ (\ref{rb-law6-oddg}) at the ground state are excited to the symmetrical positions with respect to the lines $\Im(\bar{z})= \frac{\eta}{2}$ and
$\Im(\bar{z})= \frac{\eta}{2}$, respectively.
Here, the boundary strings $w_{5e}$ contributes nothing to the energy, while the $w_{6e}$ indeed has the non-zero contribution and induce the excitation energy of the system with odd $N$.

Substituting the discrete roots (\ref{rb-law12}) and boundary strings (\ref{rb-law6-evene} ) or (\ref{rb-law6-odde}) into (\ref{rE}), we obtain the energy at the
first excited state
\begin{equation}\label{Eer}
E^{e}_{r} =e_{r}N + E^{f}_{r} +E^{+}_{r} +E^{-}_{r} +  E^{w}_{r}(w_1) +  E^{w}_{r}(w_2) +  E^{w}_{r}(w_{5e}) +  E^{w}_{r}(w_{6e}).
\end{equation}
Define the excitation energy as $\Delta E_{r}\equiv E^{e}_{r} -E^{g}_{r}$ and we have
\begin{eqnarray}
\Delta E_{r} =E^{w}_{r}(w_{5e})+E^{w}_{r}(w_{6e})-E^{w}_{r}(w_{5g})-E^{w}_{r}(w_{6g}). \label{rbE-ele}
\end{eqnarray}

\begin{figure}[htbp]
\centering
\subfigure[$N=100$]{
\includegraphics[width=3.5cm]{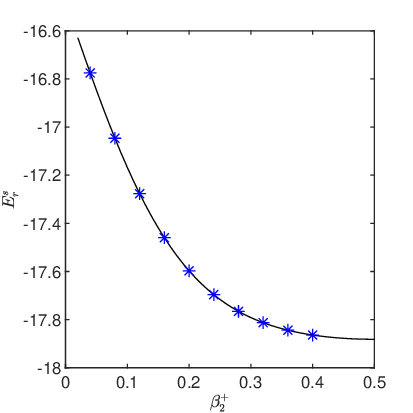}
}
\subfigure[$N=101$]{
\includegraphics[width=3.5cm]{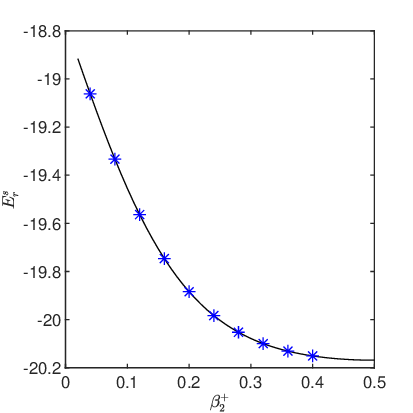}
}
\subfigure[$N=100$]{
\includegraphics[width=3.5cm]{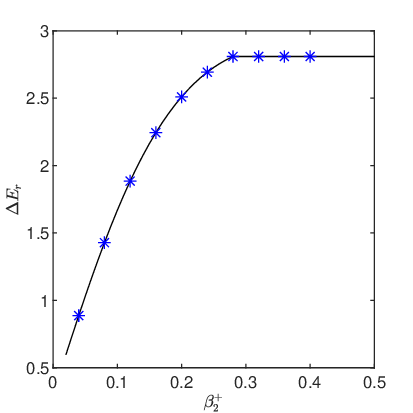}
}
\subfigure[$N=101$]{
\includegraphics[width=3.5cm]{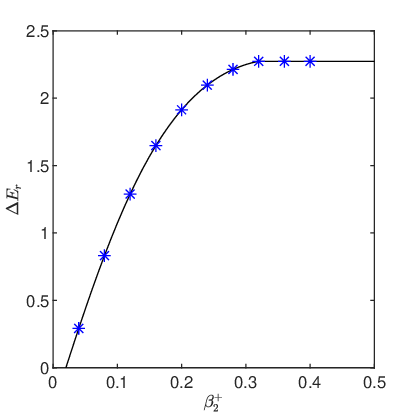}
}
\caption{
(a)-(b) The surface energy $E^s_{r}$ versus $\beta^{+}_2$. (c)-(d) The excitation energy $\Delta E_{r}$ versus the boundary parameter $\beta^{+}_2$.  Here the model parameters are chosen as $\tau=0.6i$, $\eta=0.7$,
$\beta^{+}_1=0.04$, $\beta^{-}_1=0.02$, $\beta^{-}_2=0.02$, $\beta^{+}_3=0.04i$ and $\beta^{-}_3=0.03i$.
The solid lines indicate the analytic results and the asterisks are the DMRG data. They are consistent with each other very well.
}\label{rbF-delE}
\end{figure}

We should remark that the surface energy $E^s_{r}$ and the excitation energy $\Delta E_{r}$ depend on the parity of $N$, due to the different patterns of boundary strings.
The surface energy and excitation energy versus the boundary parameter $\beta^{+}_2$ are shown in Fig.\ref{rbF-delE}.
To check the correction of our results, we also apply the density matrix renormalization group (DMRG) method \cite{white1992density} to
study $E^s_{r}$ and $\Delta E_{r}$ with several values of $\beta^{+}_2$, and the results are shown as the asterisks in Fig.\ref{rbF-delE}.
It is clear that our analytic results are consistent with the numerical ones very well.

\subsubsection{Results in the boundary parameters region (\ref{r-trans})}
\label{r-boundary}

\begin{figure}[htbp]
\centering
\subfigure[]{
\includegraphics[width=4.5cm]{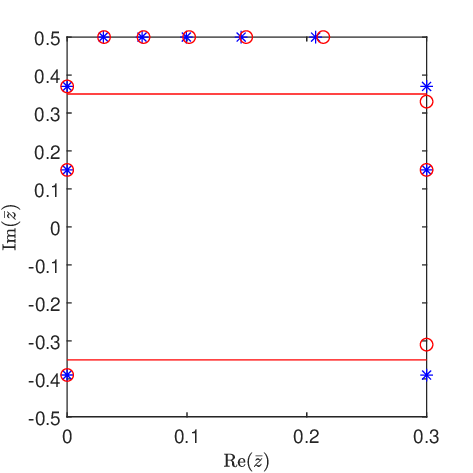}
}
\subfigure[]{
\includegraphics[width=4.5cm]{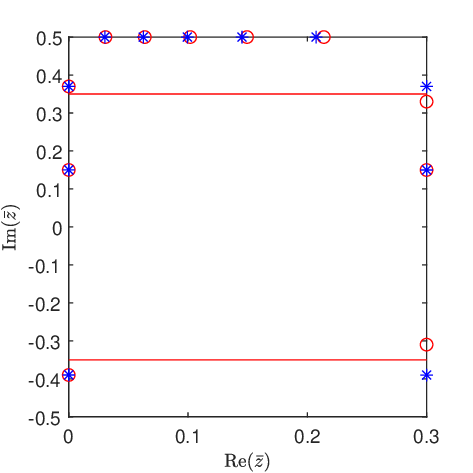}
}
\subfigure[]{
\includegraphics[width=4.5cm]{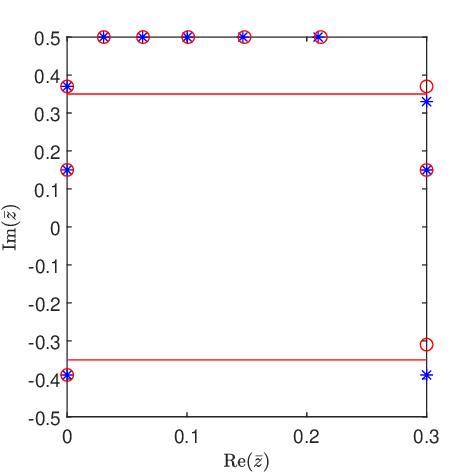}
}
\caption{Exact numerical results of the zero roots after parameter changes (a) $\beta^{+}_1\rightarrow 1-\beta^{+}_1 $, (b) $\beta^{+}_1\rightarrow \beta^{+}_1+\tau $, and (c) $\beta^{+}_1\rightarrow \beta^{+}_1+1$.
The asterisks indicate the numerical values of $\{\bar{z}\}$ at the ground state and the circles indicate the values of $\{\bar{z}\}$ at the first excited state.
Here $N=8$, $\tau=0.6i$, $\eta=0.7$, $\beta^{+}_1=0.04, \beta^{-}_1=0.02, \beta^{+}_2=0.04, \beta^{-}_2=0.02, \beta^{+}_3=0.04i$ and $\beta^{-}_3=0.03i$.
}\label{rge-tran}
\end{figure}

Next, we should investigate the remaining region of boundary parameters.
We find that the parameter changes (\ref{r-trans}) do not effect the root patterns except for the boundary strings, as shown in Fig.\ref{rge-tran} for the numerical validation.
Thus the forms of Eqs.(\ref{BAr})-(\ref{rbEroot}) keep unchanged. From the derivation, we also know that in the thermodynamic limit,
the energies $ e_{r}, E^{f}_{r}, E^{\pm}_{r}$ determined by the bulk roots are the same as before. The discrete roots only affect the values of boundary energy $E^{w}_{r}(w_t)$ (\ref{rbEroot}).
Thus we should analyze the distribution of discrete roots $\{w_t \}$.

\begin{figure}[htbp]
\centering
\subfigure[]{
\includegraphics[width=3.5cm]{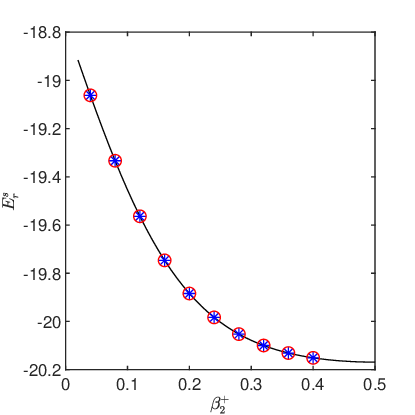}
}
\subfigure[]{
\includegraphics[width=3.5cm]{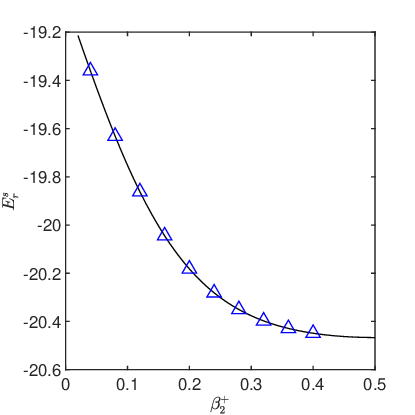}
}
\subfigure[]{
\includegraphics[width=3.5cm]{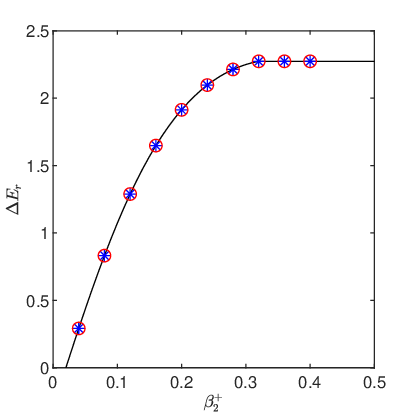}
}
\subfigure[]{
\includegraphics[width=3.5cm]{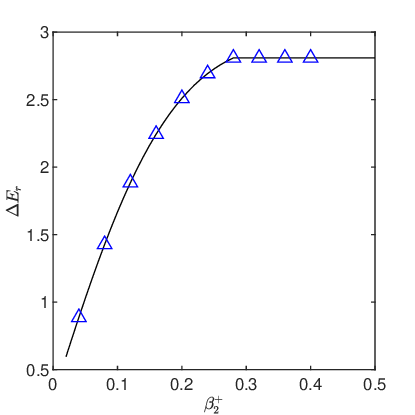}
}
\caption{
(a)-(b) The surface energy $E^s_{r}$ versus the boundary parameter $\beta^{+}_2$. (c)-(d) The excitation energy $\Delta E_{r}$ versus $\beta^{+}_2$.
The solid lines indicate the analytic results. The asterisks, circles and triangles indicate the DMRG results after the transformations $\beta^{+}_1\rightarrow 1- \beta^{+}_1$,
$\beta^{+}_1\rightarrow  \beta^{+}_1+\tau $ and $\beta^{+}_1\rightarrow \beta^{+}_1 +1$, respectively. Here $N=101, \tau=0.6i, \eta=0.7$, $\beta^{+}_1=0.04, \beta^{-}_1=0.02, \beta^{-}_2=0.02, \beta^{+}_3=0.04i$ and $\beta^{-}_3=0.03i$.
}\label{rge-verify}
\end{figure}

The numerical results show that the distribution of discrete roots is invariant under the parameter changes $\beta^{+}_1 \rightarrow 1-\beta^{+}_1$ and $\beta^{+}_1\rightarrow \beta^{+}_1+\tau $ in the ground state and the first excited state,
which can also be obtain by comparing the data in Fig.\ref{rb-gedis} (a) with those in Fig.\ref{rge-tran} (a)-(b).
Then we conclude that the surface energy $E^s_{r}$ and the excitation energy $\Delta E_{r}$ are invariant under such parameter changes in the thermodynamic limit.
We verify this conclusion by the DMRG method and the results are shown in Fig.\ref{rge-verify} (a) and (c),
where asterisks and circles are the results after taking parameter transformations, and the solid lines are
the analytic results. We see that they coincide with each other very well.

However, the parameter transformation $\beta^{+}_1\rightarrow \beta^{+}_1 +1 $ will indeed affect the distribution of discrete boundary strings at the ground state and the first excited state.
By careful checking the patterns of zero roots, we also conclude that this transition is equivalent to the change of parities of sites number $N$. For example, please compare Fig.\ref{rb-gedis}(c) with Fig.\ref{rge-tran}(c).
Thus the discrete boundary strings after taking $\beta^{+}_1\rightarrow \beta^{+}_1 +1 $ are the same as that changing the parities of $N$, which have been given in Eqs.(\ref{rb-law4})-(\ref{rb-law6-odde}).
Therefore the surface energy $E^s_{r}$ and excitation energy $\Delta E_{r}$ for the even $N$ can be obtained by using the boundary strings with the odd $N$, and vice versa.
We verify this conclusions by the DMRG method and the results are shown in Fig.\ref{rge-verify}(b) and (d),
where the triangles are the results after taking parameter transformation, and the solid lines are
the analytic results. It can be seen that the analytic results coincide with the DMRG results.

Such a parity dependence of the surface energy and the excitation energy is due to the quasi-long-range Neel order in the bulk.
For the real crossing parameter $\eta$, the coupling constants (\ref{jxyz}) satisfy $|J_x| > |J_y| >|J_z|$, which leads to the spontaneous magnetization
and the easy axis of magnetization is the $x$-direction. If the couplings are anti-ferromagnetic, two spins with nearest neighbor should take the opposite directions to reduce the energy.
For the even $N$, two boundary spins prefer to be anti-parallel. While for the odd $N$, two boundary spins prefer to be parallel.
Therefore, fixed boundary magnetic fields must induce the different surface energies and excitation energies for the different parities of $N$.
This conclusion is also valid for the ferromagnetic couplings.
Thus we explain the reason why flipping the boundary fields along the $y$- and $z$-directions (which correspond the changes of $\beta^{+}_1\rightarrow 1- \beta^{+}_1$ and $\beta^{+}_1\rightarrow  \beta^{+}_1+\tau$)
does not affect the surface and excitation energies of the system, and flipping the boundary field along the $x$-direction (which corresponds the change of $\beta^{+}_1\rightarrow  \beta^{+}_1+1$)
has the same effects to the boundary energies as that of changing the parities of $N$.

\subsection{Surface energy and excitation with $ \eta \in (0, \frac{1}{2} )$}
\label{rs}

\begin{figure}[htbp]
\centering
\subfigure[$N=8$]{
\includegraphics[width=4.5cm]{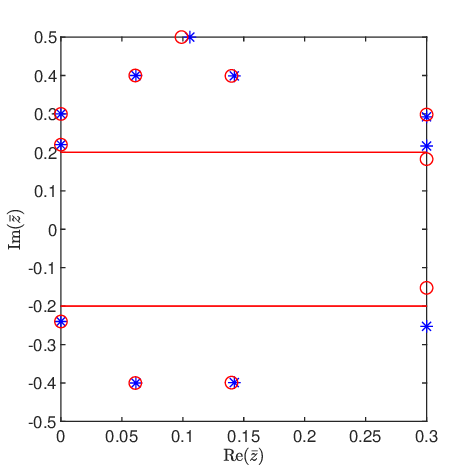}
}
\subfigure[$N=9$]{
\includegraphics[width=4.5cm]{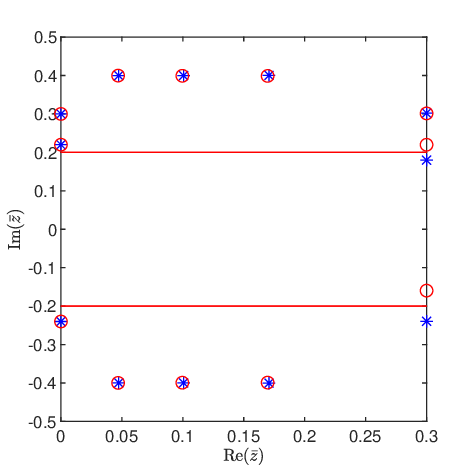}
}
\caption{
Zero roots on the complex plane at the ground state and the first excited state for (a) $N=8$ and (b) $N=9$.
The asterisks indicate the results at the ground state and the circles indicate the results at the first excited state.
For the clearness, we also draw the lines with fixed imaginary parts $\pm \frac{\eta}{2}i$.
Here $\tau=0.6i$, $\eta=0.4$, $\beta^{+}_1=0.04, \beta^{-}_1=0.02, \beta^{+}_2=0.04, \beta^{-}_2=0.02, \beta^{+}_3=0.04i$ and $\beta^{-}_3=0.03i$.
}\label{rs-gedis}
\end{figure}

If the boundary parameters are chosen in the region (\ref{r-in}) and $ 0< \eta <\frac{1}{2}$, the numerical results of zero roots at the ground state and those at the first excited state are shown in Fig.\ref{rs-gedis}.
We see that the roots include the conjugate pairs $\{\bar{z}_l=x_l \pm \eta i| l=1, \cdots, \frac{n_1}{2}\}$ with real $x_l\in [0, \frac{\tau}{2i}]$ in the bulk and
$n_2=N+3-n_1$ discrete roots $\{w_t | t=1, \cdots,n_2\}$.
The conjugate pairs in the bulk would distribute continuously in the thermodynamic limit.

Using the similar procedure as before, we obtain that the density of roots should satisfy following integration equation
\begin{eqnarray}
   &&  N \int^{\frac{\tau}{2i}}_{-\frac{\tau}{2i}}  \left( A_{\frac{\eta}{2}i}(u-x) + A_{\frac{3\eta}{2}i}(u-x)   \right) \rho(x) dx +  \sum^{n_2}_{t=1} \left( A_{w_t-\frac{\eta}{2}i}(u) + A_{w_t+\frac{\eta}{2}i}(u) \right) \nonumber \\
  &=&  A_{\eta i}(u) +A_{\eta i + \frac{i}{2}}(u ) +A_{\eta i +\frac{\tau}{2}i}(u ) +A_{\eta i +\frac{\tau-1}{2}i}(u )  - A_{\frac{\eta}{2}i}(u) - A_{\frac{\eta+1}{2}i}(u ) -A_{\frac{\eta+\tau}{2}i}(u ) \nonumber\\
   &&  -A_{\frac{\eta+1+\tau }{2}i}(u )   + N \int^{\frac{\tau}{2i}}_{-\frac{\tau}{2i}} \left\{ A_{\eta i}( u +\bar{\theta} ) +A_{\eta i}( u -\bar{\theta} ) \right\} \varrho (\bar{\theta}) d \bar{\theta}  + \sum_{\gamma=\pm} \sum^3_{l=1} A_{ \alpha^{\gamma}_l i}( u ), \label{BArs}
\end{eqnarray}
where $\rho(x)$ is the density of bulk conjugate pairs and $\varrho(\bar{\theta})$ is the density of auxiliary inhomogeneity parameters.
Taking the homogeneous limit $\varrho(\bar{\theta})\rightarrow \delta(\bar{\theta})$, we can solve the density $\rho(x)$ via Fourier transform (\ref{F}) and the result is
\begin{eqnarray}
  N \tilde{\rho}(k)  &=& \frac{1}{ \tilde{A}_{\frac{\eta}{2}i}(k) + \tilde{A}_{\frac{3\eta}{2}i}(k)} \left\{ \sum_{\gamma=\pm} \sum^3_{l=1} \tilde{A}_{ \alpha^{\gamma}_l i}( k ) + 2N \tilde{A}_{\eta i}( k )  +  \tilde{A}_{\eta i}(k) +\tilde{A}_{\eta i +\frac{i}{2}}(k ) \right. \nonumber\\
   && +\tilde{A}_{\eta i +\frac{\tau}{2}i}(k ) +\tilde{A}_{\eta i +\frac{1+\tau}{2}i}(k )  - \tilde{A}_{\frac{\eta}{2}i}(k) - \tilde{A}_{\frac{\eta+1}{2}i}(k ) -\tilde{A}_{\frac{\eta+\tau}{2}i}(k ) -\tilde{A}_{\frac{\eta+1+\tau }{2}i}(k )   \nonumber\\
   && \left. - \sum^{n_2}_{t=1} \left( \tilde{A}_{w_t-\frac{\eta}{2}i}(k) + A_{w_t+\frac{\eta}{2}i}(k) \right)    \right\} .  \label{densityF}
\end{eqnarray}
The eigen-energy in the thermodynamic limit can be expressed as
\begin{eqnarray}
  E &=&  \sum^{n_2}_{t=1} \frac{\sigma(\eta)}{\sigma'(0)} \frac{1}{2} \frac{i}{\tau} \sum^{\infty}_{k= -\infty} \left( -\tilde{B}_{ w_t-\frac{\eta}{2}i}(k) + \tilde{B}_{ w_t+\frac{\eta}{2}i}(k) \right) - N \frac{\sigma'(\eta)}{\sigma'(0)}  + \frac{\sigma'(\eta)}{\sigma'(0)}  \nonumber\\
    &&  - 2 \frac{\sigma(\eta)}{\sigma'(0)} \frac{\sigma'(2\eta)}{\sigma(2\eta)} + \frac{\sigma(\eta)}{\sigma'(0)} \frac{N}{2}  \frac{i}{\tau} \sum^{\infty}_{k= -\infty}  \left( \tilde{B}_{\frac{3\eta}{2}i}(k) - \tilde{B}_{\frac{\eta}{2}i}(k) \right) \tilde{\rho}(k). \label{rsE}
\end{eqnarray}
By substituting Eq.(\ref{densityF}) into the above Eq.(\ref{rsE}), we can ultimately solve for the energy $E$ as given by Eqs.(\ref{rE})-(\ref{rbEroot}).

Next, we should determine the discrete roots. We focus on the ground and the first excited states.
From Fig.\ref{rs-gedis}, we see that there exists a discrete root located on the line $\Im(\bar{z})=\frac{1}{2}$ which contributes nothing to the eigen-energy and only appears in the case of even $N$.
The discrete roots (\ref{rb-law12}) also exist and contribute nothing to the eigen-energy.
The remaining discrete roots include the boundary strings
located along the line $\Re(\bar{z})=0$ and the boundary strings
located along the line $\Re(\bar{z})=\frac{\tau}{2i}$. From the numerical simulation, we find that the maximum length of these boundary strings is $2\eta$.

The boundary strings $w^{-}$ and $w^{+}$ located along the line $\Re(\bar{z})=0$ are induced by the boundary parameters $\beta^{\pm}_1$.
With the increasing of boundary parameters $\beta^{-}_1$ and $\beta^{+}_1$, two discrete roots $w^{-}$ and $w^{+}$ will move away from the real axis, and eventually form the strings $w^{-}=\eta i$ and $w^{+}=-\eta i $,
which obey Eq.(\ref{string-r}) with $n_j=1, \nu_j=1$ and the distance between them is the maximum value $2\eta$. From these analyses, we present this kind of boundary strings as
\begin{eqnarray}
  w^{-} =  \phi_2 i, \quad
  w^{+} = - \min \left\{ \frac{\eta}{2} + \beta^{+}_1, 2\eta-\phi_2, \frac{1}{2} \right\} i, \label{rs-law4}
\end{eqnarray}
where $\phi_2= \min \left\{ \frac{\eta}{2} + \beta^{-}_1, \eta \right\}$.
It should be noted that as shown in Fig.\ref{rs-gedis}, these boundary strings keep unchanged at both the ground state and the first excited state, and are independent of the parities of $N$.
Clearly, this boundary strings contribute nothing to the eigen-energy of the system.

\begin{figure}[htbp]
\centering
\subfigure[$N=314$]{
\includegraphics[width=3.5cm]{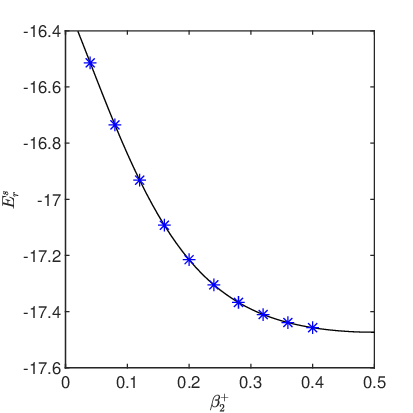}
}
\subfigure[$N=315$]{
\includegraphics[width=3.5cm]{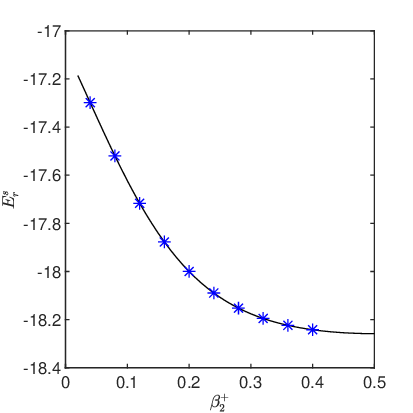}
}
\subfigure[$N=314$]{
\includegraphics[width=3.5cm]{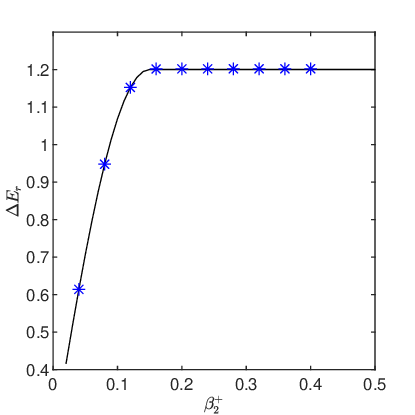}
}
\subfigure[$N=315$]{
\includegraphics[width=3.5cm]{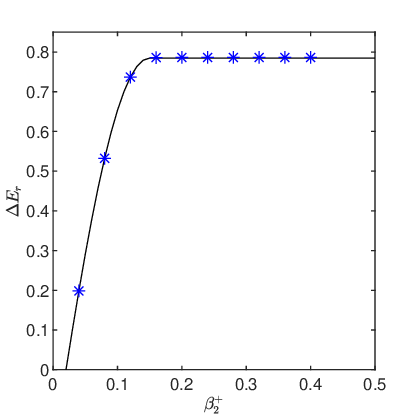}
}
\caption{The surface energy $E^s_{r}$ and the excitation energy $\Delta E_{r}$ versus the boundary parameter $\beta^{+}_2$.
The solid lines indicate the analytic results of $E^s_{r}$ calculated from Eq.(\ref{rsEs}), and $\Delta E_{r}$ calculated from Eq.(\ref{rsE-ele}).
The blue asterisks are the results obtained by using DMRG with $N=314$ [(a) and (c)] and $N=315$ [(b) and (d)].
Here, the fixed model parameters are chosen as $\tau=0.6i$, $\eta=0.3$, $\beta^{+}_1=0.04$, $ \beta^{-}_1=0.02$, $ \beta^{-}_2=0.02$, $ \beta^{+}_3=0.04i$ and $\beta^{-}_3=0.03i$.
}\label{rsF-delE}
\end{figure}

However, the boundary strings induced by the boundary parameters $\beta_{2}^{\pm}$ (the complex numbers located at the line ${\rm Re}(\bar z)=\frac{\tau}{2i}$) can contribute the non-zero value to the energy. Thus we only consider this kind of boundary strings, which depend on the states and the parities of $N$. At the ground state, the boundary strings are
\begin{eqnarray}
  w^{-}_g = \frac{\tau}{2i} + \phi_3 i, \quad
  w^{+}_g = \frac{\tau}{2i} - \min \left\{ \frac{\eta}{2} + \beta^{+}_2, 2\eta-\phi_3, \frac{1}{2} \right\} i, \label{rs-law6-eveng}
\end{eqnarray}
with $\phi_3 = \min \left\{ \frac{\eta}{2} + \beta^{-}_2, \eta \right\}$ for even $N$, and are
\begin{eqnarray}
  w^{-}_g = \frac{\tau}{2i} + \max \left\{ \frac{\eta}{2} - \beta^{-}_2, 0 \right\} i, \quad
  w^{+}_g = \frac{\tau}{2i} - \min \left\{ \frac{\eta}{2} + \beta^{+}_2, \frac{3\eta}{2}+\beta^{-}_2, \frac{1}{2} \right\} i. \label{rs-law6-oddg}
\end{eqnarray}
for odd $N$.
The ground state energy reads
\begin{equation}
E^{g}_{r} = e_{r}N + E^{f}_{r} +E^{+}_{r} +E^{-}_{r} + E^{w}_{r}(w^{-}_{g}) +  E^{w}_{r}(w^{+}_{g}).
\end{equation}
Here $e_{r}$, $E^{f}_{r}$ and $E^{\pm}_{r}$ are given by Eqs.(\ref{rbEden}), (\ref{rEfree}) and (\ref{rbEpara}), respectively.
Substituting the boundary strings (\ref{rs-law6-eveng}) or (\ref{rs-law6-oddg}) into Eq.(\ref{rbEroot}), we obtain the values of $E^{w}_{r}(w^{\pm}_{g})$.
Then the ground state energy is completely determined.
Comparing with the ground state energy of XYZ spin chain with periodic boundary condition, we obtain the surface energy of present open system as
\begin{eqnarray}
E^s_{r} = E^{f}_{r} +E^{+}_{r} +E^{-}_{r} + E^{w}_{r}(w^{-}_{g}) +  E^{w}_{r}(w^{+}_{g})- \frac{1}{2}(1- (-1)^N) E^{w}_{r}(\frac{\tau}{2i}). \label{rsEs}
\end{eqnarray}
According to this analytical expression, the surface energy $E^s_{r}$ with some fixed model parameters are shown in Fig.\ref{rsF-delE}(a) and (b) as the solid lines.
To check the correction of these results, we calculate the surface energy with same model parameters by suing the DMRG method, and the numerical data are shown in
Fig.\ref{rsF-delE}(a) and (b) as the asterisks. We see that the analytic results and numerical ones are consistent with each other very well.

Next, we consider the first excited state. By comparing the zero roots distribution (represented by asterisks) at the ground state and
the ones (represented by circles) at the first excited state in Fig.\ref{rs-gedis}, we see that the boundary strings located
along the line $\Re(\bar{z})=\frac{\tau}{2i}$ showing a different pattern.
Furthermore, we note that the boundary strings at the first excited state are dependent on the parities of $N$.

In the case of even $N$, as illustrated in subgraph (a) of Fig.\ref{rs-gedis}, it can be observed that
at the first excited state, the boundary strings $w^{-}_{g}$ and $w^{+}_{g}$ (\ref{rs-law6-eveng}) jump to their symmetrical positions with respect to the line $\Im(\bar{z})= \frac{\eta}{2}$ and $\Im(\bar{z})= -\frac{\eta}{2}$, respectively, and form the boundary strings $w^{-}_{e}$ and $w^{+}_{e}$.
The boundary strings $w^{-}_{e}$ and $w^{+}_{e}$ are located within the two red lines $\Im(\bar{z})=\pm\frac{\eta}{2}$, and they can contribute the non-zero eigen-energy.
With the increasing of boundary parameters $\beta^{-}_2$ and $\beta^{+}_2$, the discrete roots $w^{-}_{e}$ and $w^{+}_{e}$ move towards the real axis, and eventually located at the real axis.
From above analysis, we obtain the form of boundary strings as
\begin{eqnarray}
  w^{-}_{e}= \frac{\tau}{2i} + \max \left\{ \frac{\eta}{2}-\beta^{-}_2, 0 \right\} i, \quad
  w^{+}_{e}= \frac{\tau}{2i} - \max \left\{ \frac{\eta}{2}-\beta^{+}_2,  0 \right\} i. \label{rs-law6-evene}
\end{eqnarray}

In the case of odd $N$, as illustrated in subgraph (b) of Fig.\ref{rs-gedis}, it can be observed that at the first excited state,
the discrete zero roots $w^{-}_{g}$ and $w^{+}_{g}$ (\ref{rs-law6-oddg}) jump to their symmetrical positions with respect to the red lines $\Im(\bar{z})= \frac{\eta}{2}$ and $\Im(\bar{z})= -\frac{\eta}{2}$, respectively, thereby forming the discrete roots $w^{-}_{e}$ and $w^{+}_{e}$.
Furthermore, as the boundary parameters $\beta^{-}_2$ and $\beta^{+}_2$ increase, the discrete root $w^{-}_{e}$ moves away from the real axis, while the roots $w^{+}_{e}$ moves towards the real axis.
From above analysis, we obtain the form of boundary strings as
\begin{eqnarray}
  w^{-}_{e} = \frac{\tau}{2i} + \min \left\{ \frac{\eta}{2} + \beta^{-}_2, \frac{1}{2} \right\} i, \quad
  w^{+}_{e} = \frac{\tau}{2i} - \max \left\{ \frac{\eta}{2}-\beta^{+}_2, 0 \right\} i. \label{rs-law6-odde}
\end{eqnarray}
Substituting these roots into the unified form (\ref{rbEroot}), we obtain the first excited state energy
\begin{equation}
E^{e}_{r} = e_{r}N + E^{f}_{r} +E^{+}_{r} +E^{-}_{r} + E^{w}_{r}(w^{-}_{e}) +  E^{w}_{r}(w^{+}_{e}).\label{rs-1law6-odde}
\end{equation}
Subtract the result from the ground state energy, and we have the excitation energy
\begin{eqnarray}
\Delta E_{r}= E^{w}_{r}(w^{-}_{e})+E^{w}_{r}(w^{+}_{e})-E^{w}_{r}(w^{-}_{g})-E^{w}_{r}(w^{+}_{g}). \label{rsE-ele}
\end{eqnarray}
The analytical results of excitation energy with certain fixed model parameters for the different parities of $N$ are shown in Fig.\ref{rsF-delE}(c) and (d) as the solid lines,
where the asterisks are the DMRG results with the same model parameters. We see that the analytic results coincide with the DMRG data very well, and our results are correct.

For the system with $\eta\in(0,\frac{1}{2})$, due to the quasi-long-range Neel order in the bulk, it is expected that the results in the boundary parameters region (\ref{r-trans}) are consistent with the discussions presented in the subsection \ref{r-boundary}.
Firstly, the distribution of discrete roots is invariant under the parameter changes $\beta^{+}_1 \rightarrow 1-\beta^{+}_1$ and $\beta^{+}_1\rightarrow \beta^{+}_1+\tau $ in the ground state and the first excited state.
Consequently, the surface energy $E^s_{r}$ and the excitation energy $\Delta E_{r}$ remain unchanged under such parameter changes in the thermodynamic limit.

Secondly, the effect of the parameter change $\beta^{+}_1\rightarrow \beta^{+}_1 +1 $ on the discrete boundary strings is identical to the results obtained by directly changing the parity of $N$.
The specific forms of these discrete boundary strings, which can contribute a non-zero eigen-energy, have been elaborated in Eqs.(\ref{rs-law6-eveng}), (\ref{rs-law6-oddg}), (\ref{rs-law6-evene}), and (\ref{rs-law6-odde}).
Therefore the surface energy $E^s_{r}$ and excitation energy $\Delta E_{r}$ for the even $N$ can be obtained by using the boundary strings with the odd $N$, and vice versa.

\section{Thermodynamic limit for the imaginary $\eta$}
\label{sec-i}
\setcounter{equation}{0}

\subsection{Patterns of zero roots}

Next, we study the case that $\eta$ is imaginary. We first examine the distribution patterns of zero roots $\{z_l\}$.
With the help of choosing real inhomogeneity parameters $\{\theta_j \}$, we prove that $ t(u)^{\dag} =t(-u^{\ast})$, which gives $\Lambda(u) = \Lambda^{*}(-u^{*})$.
Therefore, if $z_l$ is a root of $\Lambda(u)$, there must exist another root $z_j$ satisfying
\begin{equation}\label{period-i}
z_j =z^{*}_l + m_1 \tau +m_2, \quad m_1, m_2 \in \mathbb{Z}.
\end{equation}
Based on Eq.(\ref{period-i}), we deduce that the zero roots $\{\pm z_l\}$ can be classified into:  (i) real numbers; (ii) pure imaginary numbers on the lines $\pm \frac{\tau}{2}$,
where their conjugates shifted by $\tau$ become themselves;
(iii) conjugate pairs $\{\pm z_l$, $\pm z^{\ast}_l\}$ and (iv) discrete roots induced  by the boundary magnetic fields.
According to the quasi-periodicity of elliptic functions and the relation (\ref{period-i}), we fix the roots into the region $\Im(z_l)\in [-\frac{\tau}{2i},\frac{\tau}{2i}]$ and $\Re(z_l)\in [-\frac{1}{2},\frac{1}{2}]$.

Now, we determine the detailed forms of root patterns in the cases of (iii) and (iv).
As explained previously, for the discrete values of $\eta_{L,K}$ (\ref{eta-discrete}), the eigenvalue $\Lambda(u)$ is characterized by the homogeneous $T-Q$ relation (\ref{T-Q}).
In the thermodynamic limit, the values of $\eta$ could be continuous with the properly choosing of $L$, $K$ and $M$. We should note that all the eigen-states can be
obtained by putting $M=N$, thus the energy spectrum are complete.
Putting $u_j \equiv \lambda_j -\frac{\eta}{2}$ and considering the homogeneous limit $\{\theta_j \rightarrow 0 \}$, the Bethe roots in the $T-Q$ relation (\ref{T-Q}) should satisfy the BAEs
\begin{eqnarray}
   && -e^{-8i\pi L \lambda_j} \frac{\sigma(2\lambda_j+\eta )}{\sigma(2\lambda_j-\eta )} \prod_{\gamma=\pm}\prod_{k=1}^3 \frac{\sigma(\lambda_j - \varepsilon_k^{\gamma} \alpha_{k}^{\gamma} -\frac{\eta}{2})}{\sigma(\lambda_j + \varepsilon_k^{\gamma} \alpha_{k}^{\gamma} +\frac{\eta}{2})} \left(  \frac{\sigma(\lambda_j +\frac{\eta}{2})}{\sigma(\lambda_j -\frac{\eta}{2})} \right)^{2N} \nonumber \\
   &&= \prod_{l=1}^N \frac{\sigma(\lambda_j- \lambda_l +\eta )\sigma( \lambda_j+ \lambda_l +\eta ) }{\sigma( \lambda_j- \lambda_l -\eta )\sigma(\lambda_j+ \lambda_l -\eta )}, \quad j=1,\cdots,N. \label{BAEsi}
\end{eqnarray}
The solutions of BAEs (\ref{BAEsi}) give that the Bethe roots satisfy the string hypothesis
\begin{equation}\label{string-i}
  \lambda_{j,k} = x_{j}+(\frac{n_j+1}{2} -k)\eta  +\frac{1-\nu_j}{4}\tau  +O(e^{-\delta N}), \quad 1\leq k \leq n_j,
\end{equation}
where $x_j$ is the position of the $j$-string on the real axis, $k$ means the $k$th Bethe roots in $j$-string, $O(e^{-\delta N})$ means the finite size correction,  $n_j$ is the length of $j$-string which is determined by $\frac{\eta}{\tau}$, and $\nu_j=\pm 1$
denotes the parity of $j$-string.
The center of $j$-string is the real axis if $\nu_j=1$, and is located at the line with fixed imaginary party $\frac{\tau}{2}$ in the complex plane if $\nu_j=-1$.

Substituting $\{u=z_j -\frac{\eta}{2} \}$ into the $T-Q$ relation (\ref{T-Q}), we obtain the relation between $\{z_j\}$ and $\{\lambda_j\}$ as
\begin{eqnarray}
   && -e^{-8i\pi L z_j} \frac{\sigma(2z_j+\eta )}{\sigma(2z_j-\eta )} \prod_{\gamma=\pm}\prod_{k=1}^3 \frac{\sigma(z_j - \varepsilon_k^{\gamma} \alpha_{k}^{\gamma} -\frac{\eta}{2})}{\sigma(z_j + \varepsilon_k^{\gamma} \alpha_{k}^{\gamma} +\frac{\eta}{2})} \left(  \frac{\sigma(z_j +\frac{\eta}{2})}{\sigma(z_j -\frac{\eta}{2})} \right)^{2N} \nonumber \\
   =&& \prod_{l=1}^N \frac{\sigma(z_j- \lambda_l +\eta )\sigma( z_j+ \lambda_l +\eta ) }{\sigma( z_j- \lambda_l -\eta )\sigma(z_j+ \lambda_l -\eta )}, \quad j=1,\cdots,N+3. \label{iBAE-zero}
\end{eqnarray}
The roots of functions $\Lambda(u)$ and $Q(u)$ could not be equal, i.e., $z_j \neq \lambda_j$. From the structure of Bethe roots $\{ \lambda_{j,k}\}$ (\ref{string-i}) and Eq.(\ref{iBAE-zero}), we obtain the structure of zero roots $\{z_j\}$ as
\begin{equation}
  z_{j} = x_{j} \pm \frac{n_j+1}{2}\eta  +\frac{1-\nu_j}{4}\tau +O(e^{-\delta N}).
\end{equation}
For clarity, we present two simple patterns of zero roots as below. If $n_j=1, \nu_j=1$, the root pattern is $z_{j}=x_{j} \pm \eta  $, and if $n_j=1, \nu_j=-1$, the root pattern is $z_{j}=x_{j} \pm (\frac{\tau}{2} - \eta)$ which has been shifted by $\tau$ to ensure $\Im(z_j)\in[-\frac{\tau}{2i},\frac{\tau}{2i}]$.

From the numerical simulation and singularity analysis of Eq.(\ref{iBAE-zero}), we obtain that there exist several discrete roots
\begin{equation}\label{discrete-i}
w_k^{\gamma} = \pm \left( \frac{\eta}{2} + \varepsilon_k^{\gamma} \alpha_{k}^{\gamma} \right).
\end{equation}
We also find that the distribution patterns of roots $\{ z_l| l=1,2,\cdots,N+3 \}$ in the interval $\Im(\eta) \in (0,\frac{\tau}{2i}]$ and
those in the interval $\Im(\eta) \in (\frac{\tau}{2i},\frac{\tau}{i})$ are different. Thus we should consider them separately.

\subsection{Surface energy and excitation with $ \Im(\eta)\in (\frac{\tau}{2i}, \frac{\tau}{i} )$}
\label{ib}
\begin{figure}[htbp]
\centering
\subfigure[$N=8$]{
\includegraphics[width=4.5cm]{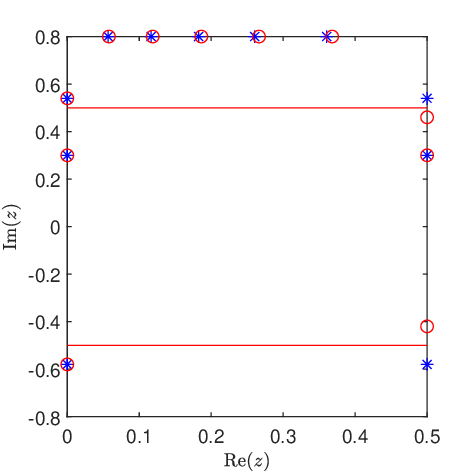}
}
\subfigure[$N=9$]{
\includegraphics[width=4.5cm]{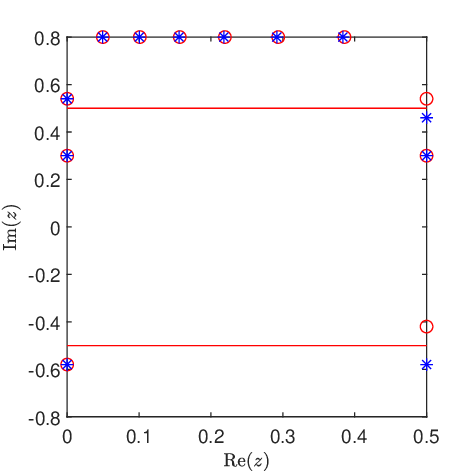}
}
\caption{ The distribution patterns of zero roots at the ground state and the first excited state for (a) $N=8$ and (b) $N=9$.
Here $\tau=1.6i$, $\eta=i$, $\beta^{+}_1=0.08i$, $ \beta^{-}_1=0.04i$, $ \beta^{+}_2=0.1$, $ \beta^{-}_2=0.04$, $ \beta^{+}_3=0.08i$ and $ \beta^{-}_3=0.04i$.
The asterisks indicate the $z$-roots at the ground state and the circles means the results at the first excited state.
The imaginary part of the red lines is $\pm \frac{\eta}{2i}$.
}\label{ib-gedis}
\end{figure}

The patterns of $z$-roots at the ground state and at the first excited state are shown in Fig.\ref{ib-gedis}.
From it, we see that the bulk of the roots are located on the line $\frac{\tau}{2}$ with the form of $ \{ z_l= x_l+\frac{\tau}{2} | l=1,\cdots,n_1 \}$ and $x_l\in [0, \frac{1}{2}]$.
These roots are continuously distributed in the thermodynamic limit. Besides, there are $n_2=N+3-n_1$ discrete roots $ \{w_t| t=1, \cdots,n_2 \}$ with the form of conjugate pairs or complex numbers located on the imaginary axis and vertical line $\Re(z)=\frac{1}{2}$.

Based on the above zero root patterns, we can study the thermodynamic limit with the suitable choice of inhomogeneity parameters $\{\theta_j \}$.
Suppose $\{\theta_j \}$ are real and define the corresponding density as $\varrho(\theta)\sim 1/N(\theta_{j}-\theta_{j-1})$.
Taking the logarithm of Eq.(\ref{identity}), making the difference of equations for real $\theta_{j}$ and $\theta_{j-1}$, and omitting the correction $O(N^{-1})$, we readily have
\begin{eqnarray}
&& \sum^{n_2}_{t=1} \left(  C_{w_t +\frac{\eta}{2}}(u) + C_{w_t -\frac{\eta}{2}}(u) \right)  + N \int^{\frac{1}{2}}_{-\frac{1}{2}}  C_{\frac{\tau - \eta}{2}}(u-x)  \rho(x)dx  \nonumber\\
&=& C_{\eta}(u) +C_{\eta+\frac{1}{2}}(u) +C_{\eta+\frac{\tau}{2}}(u) +C_{\eta+\frac{1-\tau}{2}}(u)  - C_{ \frac{\eta}{2} }(u) - C_{ \frac{\eta+1}{2}}(u)-C_{\frac{\eta + \tau}{2}}(u) \nonumber\\
   && -C_{\frac{\eta + 1+\tau}{2}}(u)  + N \int_{-\frac{1}{2}}^{\frac{1}{2}} \left( C_{\eta}( u+\theta ) + C_{\eta}( u-\theta ) \right) \varrho(\theta) d\theta + \sum_{\gamma=\pm} \sum^3_{l=1} C_{\alpha^{\gamma}_l }(u), \label{BAi}
\end{eqnarray}
where the function $\rho(x)$ is the density of bulk roots located on the line $\frac{\tau}{2}$, $C_{\xi}(u)$ is the periodic function defined in the interval $[-\frac{1}{2}, \frac{1}{2}]$ as
\begin{eqnarray}
  C_{\xi}(u) =   \frac{\sigma'( u-\xi)}{\sigma( u-\xi)} + \frac{\sigma'( u+\xi)}{\sigma( u+\xi)}, \label{Cfi}
\end{eqnarray}
and $\xi$ is a complex number with $ \Im(\xi)\in [-\frac{\tau}{i}, \frac{\tau}{i} ]$ and $ \Re(\xi) \in [-\frac{1}{2},\frac{1}{2}] $.

In the thermodynamic limit, the density of inhomogeneity parameters is the $\delta$-function, i.e., $\varrho(\theta)\rightarrow \delta(\theta)$. Then Eq.(\ref{BAi}) can be solved by the Fourier transform (\ref{F})
and the solution is
\begin{eqnarray}
    N \tilde{\rho}(k)  &=& \frac{1}{\tilde{C}_{\frac{\tau - \eta}{2}}(k) } \left\{  2 N  \tilde{C}_{\eta}( k ) + \sum_{\gamma=\pm} \sum^3_{l=1} \tilde{C}_{\alpha^{\gamma}_l }( k ) +\tilde{C}_{\eta}(k) +\tilde{C}_{\eta+\frac{1}{2}}(k) \right. \nonumber\\
   &&  +\tilde{C}_{\eta-\frac{\tau}{2}}(k)  +\tilde{C}_{\eta+\frac{1-\tau}{2}}(k) - \tilde{C}_{ \frac{\eta}{2} }(k) - \tilde{C}_{ \frac{\eta+1}{2}}(k)  -\tilde{C}_{\frac{\eta + \tau}{2}}(k) \nonumber\\
   &&\left.-\tilde{C}_{\frac{\eta + 1+\tau}{2}}(k)  - \sum^{n_2}_{t=1} \left(  \tilde{C}_{w_t +\frac{\eta}{2}}(k) + \tilde{C}_{w_t -\frac{\eta}{2}}(k) \right) \right\} , \label{ib-den}
\end{eqnarray}
where $\tilde{\rho}(k)$ is the Fourier transformation of $\rho(x)$, $\tilde{C}_{\xi}(k)$ is the Fourier transformation of $C_{\xi}(u)$
\begin{equation}\label{CfiF}
  \tilde{C}_{\xi}(k) = \left\{ \begin{array}{cc}
                      - 2i\pi   \left( - \cosh( ik\pi 2 \xi ) \coth(ik\pi \tau) + \sI(\xi) \sinh(ik\pi 2 \xi )  \right) , & k\neq 0, \\
                     0, & k=0,
                   \end{array}\right.
\end{equation}
and the function $\sI(\xi)$ is given by (\ref{sI}). Please note that the Fourier spectrum $\{k\}$ take the integer values.

Substituting the zero root patterns at the ground and first excited state into the expression of energy (\ref{Evalue}), we have
\begin{eqnarray}
  E &=&  \frac{\sigma(\eta)}{\sigma'(0)} \frac{1}{2} N \int^{\frac{1}{2}}_{-\frac{1}{2}} \left( D_{\frac{\tau - \eta}{2}} (x) -2i\pi  \right) \rho(x) dx  - N \frac{\sigma'(\eta)}{\sigma'(0)}  + \frac{\sigma'(\eta)}{\sigma'(0)}\nonumber\\
   &&- 2 \frac{\sigma(\eta)}{\sigma'(0)} \frac{\sigma'(2\eta)}{\sigma(2\eta)} + \frac{\sigma(\eta)}{\sigma'(0)} \frac{1}{2} \sum^{n_2}_{t=1} \left(  D_{w_t-\frac{\eta}{2}} (0) -D_{w_t+\frac{\eta}{2}}(0)  \right)  \nonumber\\
   &=&  \frac{\sigma(\eta)}{\sigma'(0)} \frac{1}{2} N \sum^{\infty}_{k=-\infty} \left( \tilde{D}_{\frac{\tau - \eta}{2}} (k) -2i\pi \delta_{k,0} \right) \tilde{\rho}(k)  - N \frac{\sigma'(\eta)}{\sigma'(0)}  + \frac{\sigma'(\eta)}{\sigma'(0)}  \nonumber\\
   && - 2 \frac{\sigma(\eta)}{\sigma'(0)} \frac{\sigma'(2\eta)}{\sigma(2\eta)}+ \sum^{n_2}_{t=1} \frac{\sigma(\eta)}{\sigma'(0)} \frac{1}{2} \sum^{\infty}_{k=-\infty} \left( \tilde{D}_{w_t-\frac{\eta}{2}} (k) -\tilde{D}_{w_t+\frac{\eta}{2}}(k)  \right), \label{ibE}
\end{eqnarray}
where $D_{\xi}(u)$ is the periodic function defined in the interval $[-\frac{1}{2}, \frac{1}{2}]$ as
\begin{eqnarray}
  D_{\xi}(u) =  \frac{\sigma'( u-\xi)}{\sigma( u-\xi)} - \frac{\sigma'( u+\xi)}{\sigma( u+\xi)} \label{Dfi},
\end{eqnarray}
and $\tilde{D}_{\xi}(k)$ is the Fourier transformation of $D_{\xi}(u)$ (\ref{Dfi})
\begin{equation}\label{DfiF}
  \tilde{D}_{\xi}(k) = \left\{ \begin{array}{cc}
                       2i \pi   \left( - \sinh( ik\pi 2 \xi ) \coth(ik\pi \tau) + \sI(\xi) \cosh( ik\pi 2 \xi )  \right) , & k\neq 0, \\
                     \sI(\xi) 2i\pi, & k=0.
                   \end{array}\right.
\end{equation}

Substituting Eq.(\ref{ib-den}) into Eq.(\ref{ibE}), we obtain the energy $E$ as
\begin{eqnarray}
  E = e_{i} N + E^{f}_{i} +E^{+}_{i} +E^{-}_{i} + \sum^{n_2}_{t=1} E^{w}_{i}(w_t), \label{iE}
\end{eqnarray}
where $e_{i}$ is the energy density of the bulk
\begin{eqnarray}
   e_{i} =  \frac{\sigma(\eta)}{\sigma'(0)}  \left\{ 4i\pi  \sum^{\infty}_{k= 1} \tanh (ik\pi \eta)   \frac{\cosh(ik\pi(\tau - 2\eta))}{\sinh(ik\pi \tau)}  - \frac{\sigma'(\eta)}{\sigma(\eta)}  \right\},\label{ibEden}
\end{eqnarray}
$E^{f}_{i}$ is the surface energy induced by the free open boundaries
\begin{eqnarray}
  E^{f}_{i}&=& 2i\pi \frac{\sigma(\eta)}{\sigma'(0)}  \sum^{\infty}_{k=1} \tanh(ik\pi\eta) \frac{1+\cos(k\pi )}{\sinh(ik\pi \tau)} \left\{  \cosh(ik\pi(\tau - 2\eta))  \right.  \nonumber\\
  &&  \left. + \cosh(ik\pi(\tau - i|\tau - 2\eta| )) - \cosh(ik\pi(\tau - \eta ))   - \cosh(ik\pi \eta )  \right\} \nonumber\\
  && + \frac{\sigma'(\eta)}{\sigma'(0)}  - 2 \frac{\sigma(\eta)}{\sigma'(0)} \frac{\sigma'(2\eta)}{\sigma(2\eta)} , \label{ibEfree}
\end{eqnarray}
$E^{\pm}_{i}$ indicate the contributions of two boundary fields
\begin{eqnarray}
  E^{\pm }_{i} &=& 2i\pi  \frac{\sigma(\eta)}{\sigma'(0)}  \sum^{\infty}_{k=1 }\tanh (ik\pi \eta)  \frac{1}{ \sinh(ik\pi\tau) } \left\{ \cosh(2ik\pi( \frac{\tau}{2}- \beta^{\pm }_1)) \right. \nonumber \\
  && \left. + \cosh( 2ik\pi \beta^{\pm}_2 ) + \cosh(2ik\pi( \frac{\tau}{2}- \beta^{\pm}_3 )) \cos(k\pi ) \right\} , \label{ibEpara}
\end{eqnarray}
and $E^{w}_{i}(w_t)$ is the energy induced by the discrete root $w_t$
\begin{eqnarray}
  E^{w}_{i}(w_t) = - i\pi \left(   \sI(w_t + \frac{\eta}{2}) -  \sI(w_t -\frac{\eta}{2}) \right)  \frac{\sigma(\eta)}{\sigma'(0)} \sum^{\infty}_{k= -\infty }   \frac{ \cosh( ik\pi 2w_t ) }{\cosh( ik\pi \eta )}  . \label{ibEroot}
\end{eqnarray}
According to Eq.(\ref{ibEroot}), we find that only the discrete roots located in the region of $ \Im(w_t) \in[-\frac{\eta}{2i},\frac{\eta}{2i}] $ can contribute a non-zero value to the energy.

Next, we should analyze the discrete roots. From the numerical simulation as shown in Fig.\ref{ib-gedis} ,we find that there exist two discrete zero roots
\begin{equation} \label{ib-law12}
 w_1=  \frac{\tau - \eta}{2} , \quad  w_2= \frac{1}{2} + \frac{\tau- \eta}{2},
\end{equation}
which do not depend on the boundary parameters at both the ground state and the first excited state.
According to Eq.(\ref{ibEroot}), these two roots contribute the non-zero values to the energy.

The remaining discrete roots include the boundary strings induced by parameters $\beta^{\pm}_1$ and the boundary strings induced by parameters $\beta^{\pm}_3$.
Substituting Eq.(\ref{repara}) into (\ref{discrete-i}), we obtain the boundary strings as $\frac{\eta}{2} + \varepsilon^{-}_{1} \beta^{-}_1$, $-(\frac{\eta}{2} + \varepsilon^{+}_{1} \beta^{+}_1)$,
$\frac{1}{2} + \frac{\eta}{2} + \varepsilon^{-}_{3} \beta^{-}_3$ and $\frac{1}{2} -(\frac{\eta}{2} + \varepsilon^{+}_{3} \beta^{+}_3)$.
With the increasing of boundary parameters $\beta^{\pm}_1$ and $\beta^{\pm}_3$, the values of some discrete roots change and finally are located on the line $\frac{\tau}{2}$.
From the numerical simulation, we also find that the minimum length of these conjugate pairs is $ |2\eta-\tau|$.

Based on the above analysis, we obtain the boundary strings induced by $\beta^{\pm}_1$ as
\begin{eqnarray}
  w_3 =  \minI \left\{ \frac{\eta}{2} + \beta^{-}_1, \frac{\tau}{2} \right\}, \qquad
  w_4 = - \minI \left\{ \frac{\eta}{2} + \beta^{+}_1, \frac{\tau}{2} \right\}, \label{ib-law4}
\end{eqnarray}
where the function $\minI\{r_1, \cdots,r_n \}$ means the number with the minimum imaginary part among the set $\{r_j \}$.
From Eq.(\ref{ibEroot}), we see that these boundary strings contribute nothing to the energy.
As shown in Fig.\ref{ib-gedis}, it should be noted that these boundary strings keep unchanged at both the ground state and the first excited state, and are independent of the parities of the site number $N$.

However, the boundary strings induced by parameters $\beta^{\pm}_3$ depend on the states and the parities of $N$.
At the ground state, the boundary strings read
\begin{eqnarray}
  w_{5g} = \frac{1}{2} + \minI \left\{ \frac{\eta}{2} + \beta^{-}_3, \frac{\tau}{2} \right\},\qquad
  w_{6g} = \frac{1}{2} - \minI \left\{ \frac{\eta}{2} + \beta^{+}_3, \frac{\tau}{2} \right\}, \label{ib-law6-eveng}
\end{eqnarray}
for even $N$, and
\begin{eqnarray}
  w_{5g} =\frac{1}{2} + \maxI \left\{ \frac{\eta}{2} - \beta^{-}_3, 0 \right\},\qquad
  w_{6g} = \frac{1}{2} - \minI \left\{ \frac{\eta}{2} + \beta^{+}_3, \frac{\tau}{2} \right\}, \label{ib-law6-oddg}
\end{eqnarray}
for odd $N$. Here the function $\maxI\{r_1, \cdots,r_n \}$ means the number with the maximum imaginary part among the set $\{r_j \}$.

Substituting these discrete roots into Eq.(\ref{iE}), we obtain the ground state energy
\begin{equation}
E^{g}_{i}= e_{i}N + E^{f}_{i} +E^{+}_{i} +E^{-}_{i} +  E^{w}_{i}(w_1) +  E^{w}_{i}(w_2) +  E^{w}_{i}(w_{5g}) +  E^{w}_{i}(w_{6g}).
\end{equation}
The surface energy of the system with open boundary condition is defined as $E^s_{i} = E^{g}_{i}- E^{gp}_{i} $,
where $E^{gp}_{i}$ is the ground state energy of the periodic XYZ spin chain with the imaginary $\eta$ and $E^{gp}_{i}= Ne_{i} +\frac{1}{2}(1- (-1)^N) E^i(\frac{1}{2}) $.
Then we obtain the surface energy of the system as
\begin{eqnarray}
E^s_{i} &=& E^{f}_{i} +E^{+}_{i} +E^{-}_{i} +  E^{w}_{i}(w_1) +  E^{w}_{i}(w_2) +  E^{w}_{i}(w_{5g}) \nonumber\\
 &&  +  E^{w}_{i}(w_{6g})- \frac{1}{2}(1- (-1)^N) E^i(\frac{1}{2}). \label{ibEs}
\end{eqnarray}

Now, we consider the first excited state.
From Fig.\ref{ib-gedis}, we observe that the discrete roots in the ground state (represented by asterisks) are consistent with those in the first excited state (represented by circles),
except that the boundary strings located on the vertical line $\Re(z)=\frac{1}{2}$ showing a different pattern. Therefore, the first excitation is due to the redistribution of the boundary strings located on the line $\Re(z)=\frac{1}{2}$.
Furthermore, we note that the boundary strings at the first excited state are dependent on the parity of $N$.

In the case of even $N$, as illustrated in subgraph (a) of Fig.\ref{ib-gedis}, it can be observed that in the transformation from the ground state to the first excited state,
the discrete zero roots $w_{5g}$ and $w_{6g}$ (\ref{ib-law6-eveng}), which form the boundary strings located on the vertical line $\Re(z)=\frac{1}{2}$ at the ground state,
jump to their symmetrical positions with respect to the lines $\Im(z)=\frac{\eta}{2i}$ and $\Im(z)=-\frac{\eta}{2i}$, respectively, thereby forming new boundary strings $w_{5e}$ and $w_{6e}$ at the first excited state.
The discrete roots $w_{5e}$ and $w_{6e}$ are located within the two red lines $\Im(z)=\pm\frac{\eta}{2i}$, and they can contribute the non-zero excitation energy.
Additionally, as the boundary parameters $|\beta^{-}_3|$ and $|\beta^{+}_3|$ increase, the two discrete roots $w_{5e}$ and $w_{6e}$ will move towards the real number axis, and eventually form the
boundary strings $w_{5e}=\frac{1}{2}+\eta-\frac{\tau}{2}$ and $w_{6e}=\frac{1}{2}-(\eta-\frac{\tau}{2}) $ which obey Eq.(\ref{string-i}) with $n_j=1, \nu_j=-1$.
At the same time, the distance between the discrete roots is the minimum value $|2\eta-\tau|$.
Therefore, at the first excited state, the boundary string located on the vertical line $\Re(z)=\frac{1}{2}$  are
\begin{eqnarray}
 && w_{5e} = \frac{1}{2} + \psi_1 , \quad  w_{6e} = \frac{1}{2} - \maxI \left\{ \frac{\eta}{2}-\beta^{+}_3, 2\eta -\tau -\psi_1 , 0 \right\}, \label{ib-law6-evene}
\end{eqnarray}
where $ \psi_1 = \maxI \left\{ \frac{\eta}{2}-\beta^{-}_3, \eta-\frac{\tau}{2} \right\} $.

In the case of odd $N$, as illustrated in subgraph (b) of Fig.\ref{ib-gedis}, it can be observed that in the transformation from the ground state to the first excited state, the discrete zero roots $w_{5g}$ and $w_{6g}$ (\ref{ib-law6-oddg})
jump to their symmetrical positions with respect to the lines $\Im(z)= \frac{\eta}{2i}$ and $\Im(z)= -\frac{\eta}{2i}$, respectively, thereby forming new discrete roots $w_{5e}$ and $w_{6e}$.
Furthermore, as the boundary parameters $|\beta^{-}_3|$ and $|\beta^{+}_3|$ increase, the discrete root $w_{5e}$ moves away from the real axis, while $w_{6e}$ moves towards the real axis.
From above analysis, we conclude that at the first excited state, the boundary strings are
\begin{eqnarray}
  w_{5e}= \frac{1}{2} + \minI \left\{ \frac{\eta}{2} + \beta^{-}_3, \frac{\tau}{2} \right\},\quad
  w_{6e}= \frac{1}{2} - \maxI \left\{ \frac{\eta}{2}-\beta^{+}_3, \frac{3\eta}{2}-\tau -\beta^{-}_3, 0 \right\}. \label{ib-law6-odde}
\end{eqnarray}

Substituting the discrete roots into Eq.(\ref{iE}), we obtain the eigen-energy at the first excited state
\begin{equation}
E^{e}_{i}=e_{i}N + E^{f}_{i} +E^{+}_{i} +E^{-}_{i} +  E^{w}_{i}(w_1) +  E^{w}_{i}(w_2) +  E^{w}_{i}(w_{5e}) +  E^{w}_{i}(w_{6e}).
\end{equation}
The excitation energy $\Delta E_{i} = E^{e}_{i}-E^{g}_{i}$ is
\begin{eqnarray}
\Delta E_{i}= E^{w}_{i}(w_{5e})+E^{w}_{i}(w_{6e})-E^{w}_{i}(w_{5g})-E^{w}_{i}(w_{6g}). \label{ibE-ele}
\end{eqnarray}

\begin{figure}[htbp]
\centering
\subfigure[$N=100$]{
\includegraphics[width=3.5cm]{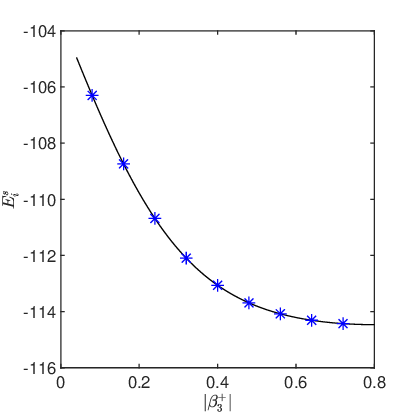}
}
\subfigure[$N=101$]{
\includegraphics[width=3.5cm]{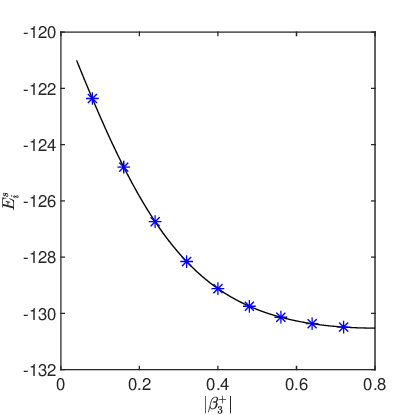}
}
\subfigure[$N=100$]{
\includegraphics[width=3.5cm]{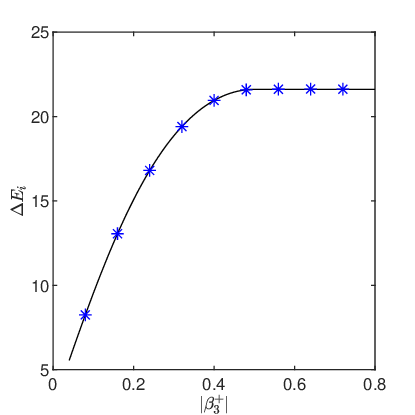}
}
\subfigure[$N=101$]{
\includegraphics[width=3.5cm]{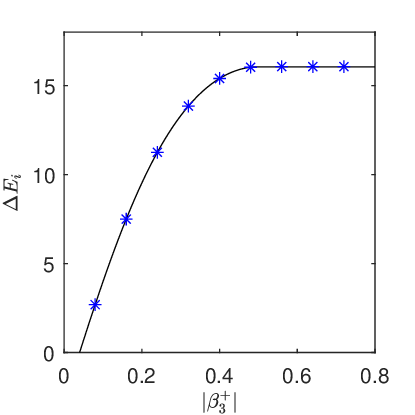}
}
\caption{
(a)-(b) The surface energy $E^s_{i}$ versus the $|\beta^{+}_3|$. (c)-(d) The excitation energy $\Delta E_{i}$ versus the boundary parameter $|\beta^{+}_3|$.
The solid lines indicate the analytic results and the asterisks are the ones obtained via DMRG with $N=100$ and $N=101$.
Here the model parameters are chosen as $\tau=1.6i$, $\eta=i$, $\beta^{+}_1=0.08i$, $\beta^{-}_1=0.04i$, $\beta^{+}_2=0.1$,
$\beta^{-}_2=0.04$ and $\beta^{-}_3=0.04i$.
}\label{ibF-delE}
\end{figure}

We remark that both the surface energy $E^s_{i}$ and the excitation energy $\Delta E_{i}$ depend on the parities of $N$.
In order to check our calculation, we also compute the surface energy and the excitation energy by using the DMRG method and the results are shown in Fig.\ref{ibF-delE}.
From it, we see that the analytic results coincide with the numerical ones very well.

\subsection{Surface energy and excitation with $ \Im(\eta) \in (0, \frac{\tau}{2i} )$}
\label{is}
\begin{figure}[htbp]
\centering
\subfigure[$N=8$]{
\includegraphics[width=4.5cm]{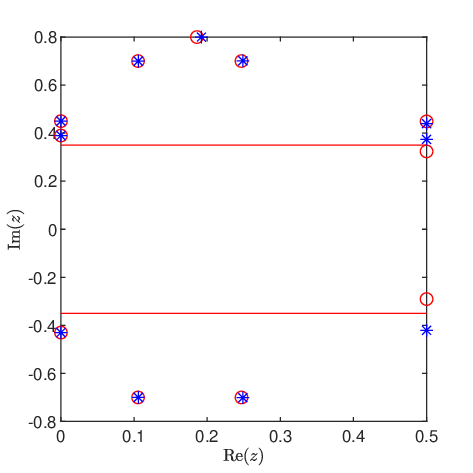}
}
\subfigure[$N=9$]{
\includegraphics[width=4.5cm]{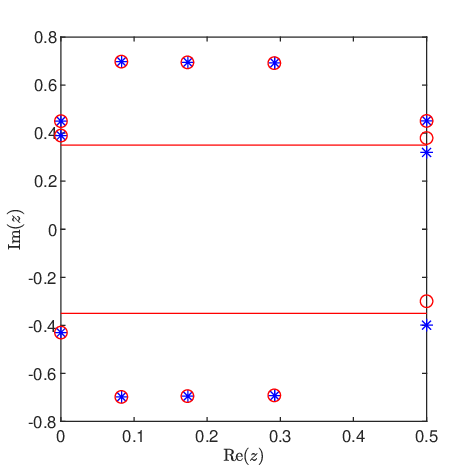}
}
\caption{The distributions of $z$-roots on the complex plane at the ground state (asterisks) and those at the first excited state (circles) for (a) $N=8$ and (b) $N=9$.
Here $\tau=1.6i$, $\eta=0.7i$, $\beta^{+}_1=0.08i$, $ \beta^{-}_1=0.04i$, $ \beta^{+}_2=0.1$, $ \beta^{-}_2=0.04$, $ \beta^{+}_3=0.05i$ and $ \beta^{-}_3=0.03i$.
The two sold lines characterize the region from $- \frac{\eta}{2}$ to $\frac{\eta}{2}$.
}\label{is-gedis}
\end{figure}

Now, we consider the region $\Im(\eta) \in (0,\frac{\tau}{2i})$. The distribution patterns of the zero roots at the ground state and the first excited state are shown in Fig.\ref{is-gedis}.
From it, we see that the roots in the bulk form the conjugate pairs $\{z_l=x_l \pm \eta | l=1,\cdots, \frac{n_1}{2}\}$ with real $x_l\in [0, \frac{1}{2}]$.
Besides, there are also $n_2=N+3-n_1$ discrete roots $\{w_t| t=1,\cdots,n_2\}$, which are real or complex.

Following the same procedure as in the previous subsection \ref{ib}, substituting the above root patterns into Eq.(\ref{identity}) and taking the thermodynamic limit, we find that the
density $\rho(x)$ of the zero roots should satisfy the constraint
\begin{eqnarray}
&& N \int^{\frac{1}{2}}_{-\frac{1}{2}} \left( C_{ \frac{3\eta}{2} }(u-x)  + C_{ \frac{\eta}{2} }(u-x)  \right) \rho(x) dx  +  \sum^{n_2}_{t=1} \left( C_{w_t+\frac{\eta}{2}}(u) + C_{w_t-\frac{\eta}{2}}(u) \right)   \nonumber\\
&=& C_{\eta}(u) +C_{\eta+\frac{1}{2}}(u) +C_{\eta+\frac{\tau}{2}}(u) +C_{\eta+\frac{1-\tau}{2}}(u) - C_{ \frac{\eta}{2} }(u) - C_{ \frac{\eta+1}{2}}(u) -C_{\frac{\eta + \tau}{2}}(u)\nonumber\\
   &&  -C_{\frac{\eta + 1+\tau}{2}}(u) + N \int_{-\frac{1}{2}}^{\frac{1}{2}} \left( C_{\eta}( u+\theta ) + C_{\eta}( u-\theta ) \right) \varrho(\theta) d\theta + \sum_{\gamma=\pm} \sum^3_{l=1} C_{\alpha^{\gamma}_l }(u), \label{BAis}
\end{eqnarray}
where $\varrho(\theta)$ is the density of inhomogeneous parameters.
In the thermodynamic limit, the density $\varrho(\theta)$ becomes the $\delta$-function, $\varrho(\theta)\rightarrow \delta(\theta)$.
Solving Eq.(\ref{BAis}) by the Fourier transform, we obtain
\begin{eqnarray}
 N  \tilde{\rho}(k) &=& \frac{1}{ \tilde{C}_{  \frac{3\eta}{2} }(k)  + \tilde{C}_{ \frac{\eta}{2} }(k) } \left\{ 2 N  \tilde{C}_{\eta}(k )  + \sum_{\gamma=\pm} \sum^3_{l=1} \tilde{C}_{ \alpha^{\gamma}_l}(k)  + \tilde{C}_{\eta}(k) + \tilde{C}_{\eta+\frac{1}{2}}(k) \right.\nonumber\\
&&  + \tilde{C}_{\eta+\frac{\tau }{2}}(k) + \tilde{C}_{\eta+\frac{1+ \tau }{2}}(k)   - \tilde{C}_{\frac{\eta}{2}}(k) - \tilde{C}_{\frac{\eta+1 }{2}}(k) - \tilde{C}_{\frac{\eta +\tau }{2}}(k) - \tilde{C}_{\frac{\eta+1+\tau }{2}}(k)  \nonumber\\
 && \left. -  \sum^{n_2}_{t=1} \left( \tilde{C}_{w_t+\frac{\eta}{2}}( k ) + \tilde{C}_{w_t-\frac{\eta}{2}}( k ) \right)  \right\} .\label{den-si}
\end{eqnarray}

According to the root patterns, the energy (\ref{Evalue}) in the thermodynamic limit can be expressed as
\begin{eqnarray}
   E &=& \frac{\sigma(\eta)}{\sigma'(0)} \frac{1}{2} N \sum^{\infty}_{k=- \infty} \left(  \tilde{D}_{\frac{\eta}{2}}(k) -\tilde{D}_{\frac{3 \eta}{2}}(k)   \right) \tilde{\rho}(k) - N \frac{\sigma'(\eta)}{\sigma'(0)}  + \frac{\sigma'(\eta)}{\sigma'(0)} \nonumber\\
    && - 2 \frac{\sigma(\eta)}{\sigma'(0)} \frac{\sigma'(2\eta)}{\sigma(2\eta)}+ \sum^{n_2}_{t=1} \frac{\sigma(\eta)}{\sigma'(0)} \frac{1}{2} \sum^{\infty}_{k=- \infty} \left( \tilde{D}_{w_t-\frac{\eta}{2}}(k) -\tilde{D}_{w_t+\frac{\eta}{2}}(k)  \right). \label{isE}
\end{eqnarray}
Substituting Eq.(\ref{den-si}) into (\ref{isE}), we find that the energy can still be expressed as (\ref{iE}).

The next task is to analyze the discrete roots. From Fig.\ref{is-gedis}, one can see that there exists a discrete root located on the line $\Im(z)=\frac{\tau}{2i}$ which contribute nothing to the eigen-energy of the system and only appears in the case of even $N$.
The discrete roots Eq.(\ref{ib-law12}) also exist and contribute nothing to the energy.
The remaining discrete roots include the boundary strings located along the line $\Re(z)=0$ and the boundary strings located along the line $\Re(z)=\frac{1}{2}$.
From the numerical simulation, we find that the maximum length of these boundary strings is $|2\eta|$.

The boundary strings located along the line $\Re(z)=0$ are induced the boundary parameters $\beta^{\pm}_1$. We denote them as $w^{-}$ and $w^{+}$.
As the boundary parameters $|\beta^{-}_1|$ and $|\beta^{+}_1|$ increase,
the two discrete roots $w^{-}$ and $w^{+}$ will move further away from the real axis, and eventually form the strings $w^{-}=\eta$ and $w^{+}=-\eta $ which obey Eq.(\ref{string-i}) with $n_j=1, \nu_j=1$.
At the same time, the distance between the discrete roots is the maximum value $|2\eta|$.
Based on the above analysis, we present the boundary strings as
\begin{eqnarray}
  w^{-} =  \psi_2, \qquad
  w^{+} = - \minI \left\{ \frac{\eta}{2} + \beta^{+}_1, 2\eta- \psi_2, \frac{\tau}{2} \right\}, \label{is-law4}
\end{eqnarray}
where $\psi_2= \minI \left\{ \frac{\eta}{2} + \beta^{-}_1, \eta \right\}$.
It should be noted that as shown in Fig.\ref{is-gedis}, these boundary strings keep unchanged at both the ground state and the first excited state, and are independent of the parities of $N$.
Clearly, this boundary strings contribute nothing to the energy of the system.

However, the boundary strings induced by the $\beta^{\pm}_3$
can contribute the non-zero values to the energy, because they are located at the region between two red lines with fixed imaginary part $\pm \frac{\eta}{2}$.
Further analysis gives that these boundary strings depend on the states as well as the parities of $N$.
With the changing of $\beta^{\pm}_3$, the maximum length of these boundary strings is $|2\eta|$.

At the ground state, the boundary strings are
\begin{eqnarray}
 && w^{-}_{g} = \frac{1}{2} + \psi_3 , \quad
  w^{+}_{g} = \frac{1}{2} - \minI \left\{ \frac{\eta}{2} + \beta^{+}_3, 2\eta-\psi_3, \frac{\tau}{2} \right\}, \label{is-law6-eveng}
\end{eqnarray}
where $ \psi_3 = \minI \left\{ \frac{\eta}{2} + \beta^{-}_3, \eta \right\} $ for the even $N$, and are
\begin{eqnarray}
  w^{-}_{g} = \frac{1}{2} + \maxI \left\{ \frac{\eta}{2} - \beta^{-}_3, 0 \right\},\quad
  w^{+}_{g} = \frac{1}{2} - \minI \left\{ \frac{\eta}{2} + \beta^{+}_3, \frac{3\eta}{2}+\beta^{-}_3, \frac{\tau}{2} \right\}, \label{is-law6-oddg}
\end{eqnarray}
for the odd $N$. Substituting solutions (\ref{is-law6-eveng})-(\ref{is-law6-oddg}) into (\ref{iE}), we obtain the ground state energy
\begin{equation}
E^{g}_{i}= e_{i}N + E^{f}_{i} +E^{+}_{i} +E^{-}_{i} +  E^{w}_{i}(w^{-}_{g})  +  E^{w}_{i}(w^{+}_{g}).
\end{equation}
The surface energy of the system is
\begin{eqnarray}
E^s_i = E^{f}_{i} +E^{+}_{i} +E^{-}_{i} +  E^{w}_{i}(w^{-}_{g})  +  E^{w}_{i}(w^{+}_{g})- \frac{1}{2}(1- (-1)^N) E^i(\frac{1}{2}). \label{isEs}
\end{eqnarray}

Now, we consider the first excited state. By comparing the distribution of zero roots (represented by asterisks) at the ground state and
the ones (represented by circles) at the first excited state in Fig.\ref{is-gedis}, it can be observed that the boundary strings located along the line $\Re(z)=\frac{1}{2}$
show a different pattern. Furthermore, we note that the boundary strings at the first excited state are dependent on the parity of $N$.

In the case of even $N$, as illustrated in subgraph (a) of Fig.\ref{is-gedis}, it can be seen that at the first excited state,
the discrete roots $w^{-}_{g}$ and $w^{+}_{g}$ (\ref{is-law6-eveng}) at the ground state jump to their symmetrical positions with respect to the lines $\Im(z)= \frac{\eta}{2i}$ and $\Im(z)=- \frac{\eta}{2i}$, respectively,
thereby forming new boundary strings $w^{-}_{e}$ and $w^{+}_{e}$ in the first excited state.
The boundary strings $w^{-}_{e}$ and $w^{+}_{e}$ are located within the two red lines $\Im(z)=\pm\frac{\eta}{2i}$, and have the contributions to the eigen-energy.
Additionally, as the boundary parameters $|\beta^{-}_3|$ and $|\beta^{+}_3|$ increase, two discrete roots $w^{-}_{e}$ and $w^{+}_{e}$ will move towards the real axis, and eventually located on the real axis.
Based on the above analysis, we obtain the boundary strings at the first excited state located on the vertical line $\Re(z)=\frac{1}{2}$ as
\begin{eqnarray}
  w^{-}_{e} = \frac{1}{2} + \maxI \left\{ \frac{\eta}{2}-\beta^{-}_3, 0 \right\}, \quad
  w^{+}_{e} = \frac{1}{2} - \maxI \left\{ \frac{\eta}{2}-\beta^{+}_3,  0 \right\}. \label{is-law6-evene}
\end{eqnarray}

In the case of odd $N$, as illustrated in subgraph (b) of Fig.\ref{is-gedis}, it can be observed that at the first excited state,
the discrete zero roots $w^{-}_{g}$ and $w^{+}_{g}$ (\ref{is-law6-oddg}) at the ground state jump to their symmetrical positions with respect to the lines $\Im(z)= \frac{\eta}{2i}$ and $\Im(z)=- \frac{\eta}{2i}$, respectively,
thereby forming new discrete roots $w^{-}_{e}$ and $w^{+}_{e}$.
Furthermore, as the boundary parameters $|\beta^{-}_3|$ and $|\beta^{+}_3|$ increase, the discrete root $w^{-}_{e}$ moves away from the real axis, while the $w^{+}_{e}$ moves towards the real axis.
Therefore, at the first excited state, the boundary strings are
\begin{eqnarray}
  w^{-}_{e} = \frac{1}{2} + \minI \left\{ \frac{\eta}{2} + \beta^{-}_3, \frac{\tau}{2} \right\}, \quad
  w^{+}_{e} = \frac{1}{2} - \maxI \left\{ \frac{\eta}{2}-\beta^{+}_3, 0 \right\}.\label{is-law6-odde}
\end{eqnarray}

Substituting solutions (\ref{is-law6-evene})-(\ref{is-law6-odde}) into (\ref{iE}), we obtain the energy at first excited state
\begin{equation}
E^{e}_{i}=e_{i}N + E^{f}_{i} +E^{+}_{i} +E^{-}_{i} +  E^{w}_{i}(w^{-}_{e})  +  E^{w}_{i}(w^{+}_{e}).
\end{equation}
The excitation energy is
\begin{eqnarray}
\Delta E_i &=& E^{w}_{i}(w^{-}_{e})+E^{w}_{i}(w^{+}_{e})-E^{w}_{i}(w^{-}_{g})-E^{w}_{i}(w^{+}_{g}). \label{isE-ele}
\end{eqnarray}

\begin{figure}[htbp]
\centering
\subfigure[$N=314$]{
\includegraphics[width=3.5cm]{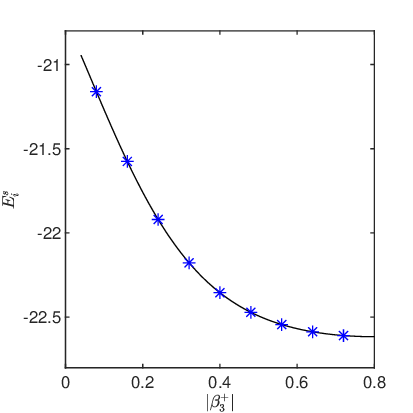}
}
\subfigure[$N=315$]{
\includegraphics[width=3.5cm]{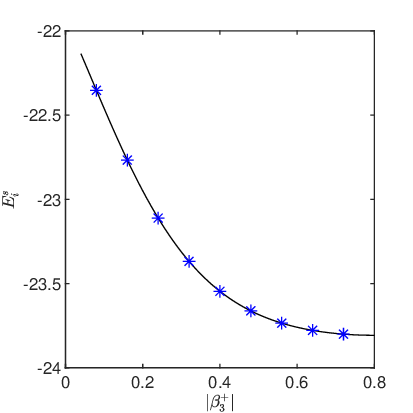}
}
\subfigure[$N=314$]{
\includegraphics[width=3.5cm]{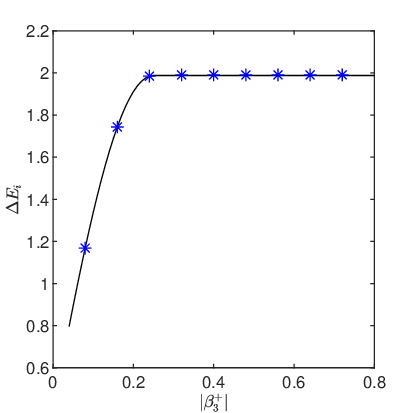}
}
\subfigure[$N=315$]{
\includegraphics[width=3.5cm]{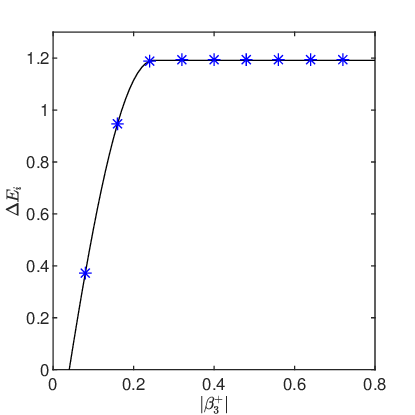}
}
\caption{
(a)-(b) The surface energy $E^s_i$ versus the $|\beta^{+}_3|$. (c)-(d) The excitation energy $\Delta E_i$ versus the boundary parameter $|\beta^{+}_3|$.
The solid lines indicate the analytic results and the asterisks are the ones obtained via DMRG method for $N=314$ and $N=315$.
Here the model parameters are chosen as $\tau=1.6i$, $\eta=0.5i$, $\beta^{+}_1=0.08i$, $\beta^{-}_1=0.04i$, $ \beta^{+}_2=0.1$, $ \beta^{-}_2=0.04$ and $\beta^{-}_3=0.04i$.
}\label{isF-delE}
\end{figure}

Now we check above analytic results by the numerical calculation. The surface energy $E^s_{i}$ and the excitation energy $\Delta E_{i}$
computed by the DMRG with $N=314$ and $N=315$ are shown in Fig.\ref{isF-delE} as the asterisks, where the solid lines are the analytic results.
We see that they are coincide with each other very well.

\subsection{Results in the boundary parameters region (\ref{i-trans})}
\label{i-boundary}

\begin{figure}[htbp]
\centering
\subfigure[]{
\includegraphics[width=4.5cm]{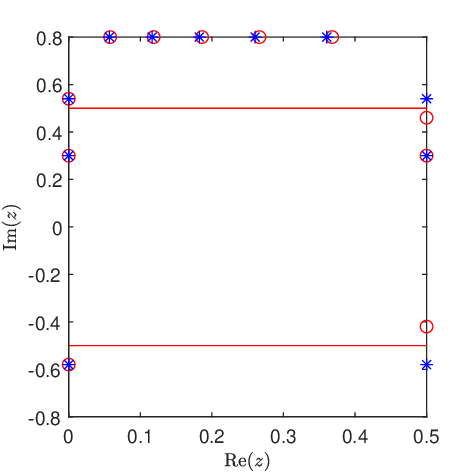}
}
\subfigure[]{
\includegraphics[width=4.5cm]{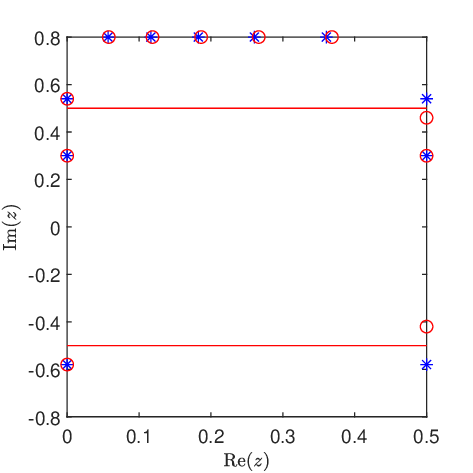}
}
\subfigure[]{
\includegraphics[width=4.5cm]{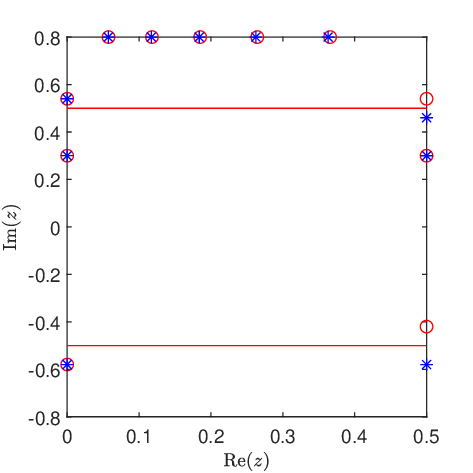}
}
\caption{ The distributions of $z$-roots at the ground state (asterisks) and those at the first excited state (circles) after the parameter changing (a) $\beta^{+}_1\rightarrow \tau-\beta^{+}_1 $, (b) $\beta^{+}_1\rightarrow \beta^{+}_1+1 $, and (c) $\beta^{+}_1\rightarrow \beta^{+}_1+\tau$.
Here $N=8$, $\tau=1.6i$, $\eta=i$, $\beta^{+}_1=0.08i$, $ \beta^{-}_1=0.04i$, $ \beta^{+}_2=0.1$, $ \beta^{-}_2=0.04$, $ \beta^{+}_3=0.08i$ and $ \beta^{-}_3=0.04i$.
}\label{ige-tran}
\end{figure}

To completely quantify the contribution of the boundary fields, it is necessary to further investigate the impact of parameter changes (\ref{i-trans}) on the distribution of the zero roots.
The parameter changes (\ref{i-trans}) do not change the root patterns except for the boundary strings, as shown in Fig.\ref{ige-tran} for the numerical validation.
Notably, the integrable equation (\ref{BAi}) satisfied by the density of roots are the same under such changes, which gives that the energies $ e_{i}, E^{f}_{i}$ and $E^{\pm}_{i}$ derived from Eqs.(\ref{ibEden}), (\ref{ibEfree}) and (\ref{ibEpara}) keep unchanged.
Therefore, we need only consider the effects of changes (\ref{i-trans}) on the discrete roots $\{w_t \}$ thus the energy $E^{w}_{i}(w_t)$ (\ref{ibEroot}).

\begin{figure}[htbp]
\centering
\subfigure[]{
\includegraphics[width=3.5cm]{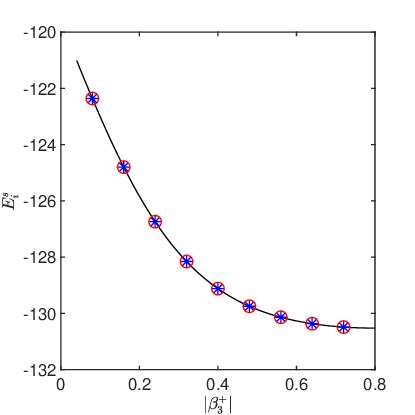}
}
\subfigure[]{
\includegraphics[width=3.5cm]{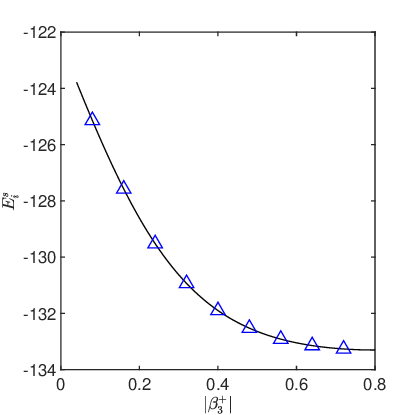}
}
\subfigure[]{
\includegraphics[width=3.5cm]{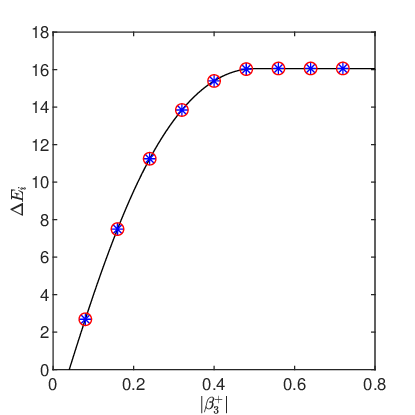}
}
\subfigure[]{
\includegraphics[width=3.5cm]{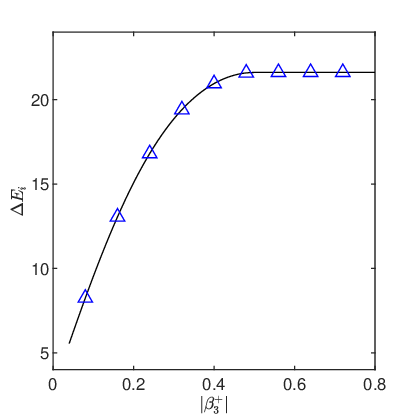}
}
\caption{The surface energy $E^s_i$ [(a)-(b)] and the excitation energy $\Delta E_i$ [(c)-(d)]
versus the boundary parameter $|\beta^{+}_3|$. The solid lines indicate the analytic results. The asterisks, circles and triangles
are the DMRG results with $\beta^{+}_1\rightarrow \tau - \beta^{+}_1$, $\beta^{+}_1\rightarrow  \beta^{+}_1+1 $ and $\beta^{+}_1\rightarrow \beta^{+}_1 +\tau$, respectively.
Here $N=101$, $\tau=1.6i$, $ \eta=i$, $\beta^{+}_1=0.08i$, $ \beta^{-}_1=0.04i$, $ \beta^{+}_2=0.1$, $ \beta^{-}_2=0.04$ and $ \beta^{-}_3=0.04i$.
}\label{ige-verify}
\end{figure}

The distributions of discrete roots are the same as before under the parameter changes $\beta^{+}_1 \rightarrow \tau-\beta^{+}_1$ and $\beta^{+}_1\rightarrow \beta^{+}_1+1 $ at the ground state and the first excited state.
This view is also supported by comparing Fig.\ref{ib-gedis} (a) and Fig.\ref{ige-tran} (a), (b).
Therefore, we conclude that the surface energy $E^s_{i}$ and excitation energy $\Delta E_{i}$ keep unchanged under such parameter changes.
We verify above conclusion by using the DMRG and the results are shown in Fig.\ref{ige-verify} (a) and (c).
It is clear that the analytic results coincide with the numerical ones very well.

However, the parameter change $\beta^{+}_1\rightarrow \beta^{+}_1 +\tau $ would affect the distribution of boundary strings at the ground state and the first excited state.
We find that the boundary strings after taking the transition $\beta^{+}_1\rightarrow \beta^{+}_1 +\tau $ are the same as that with the changing of the parities of $N$.
Therefore, under the parameter change, the surface energy $E^s_{i}$ and excitation energy $\Delta E_{i}$ for the even site number can be obtained
by using the boundary strings with odd site number, and vice versa.
We verify this conclusion with DMRG and the results are shown in Fig.\ref{ige-verify} (b) and (d).
Again, the analytic results and numerical ones coincide with each other very well.

Such a parity dependence of the surface energy and the excitation energy is due to the long-range Neel order in the bulk.
For the imaginary crossing parameter $\eta$, the coupling constant (\ref{jxyz}) gives $|J_x| < |J_y| <|J_z|$.
This leads to the spontaneous magnetization and the easy-axis is the $z$-axis. If the couplings are anti-ferromagnetic,
two boundary spins prefer to be anti-parallel along the $z$-direction for the even $N$, while two boundary spins prefer to be parallel for the odd $N$.
Therefore, the fixed boundary fields must induce the different surface energies and excitation energy for the different parities of $N$.
This also is the reason why flipping the boundary field along the $x$- and $y$-axis (which occur if we taking the transformations $\beta^{+}_1\rightarrow \tau- \beta^{+}_1$
and $\beta^{+}_1\rightarrow  \beta^{+}_1+1$) do not affect on the energies,
while flipping the boundary field along the $z$-axis (which occurs with the changing $\beta^{+}_1\rightarrow  \beta^{+}_1+\tau$)
has the same effect as that with the changing of parities of $N$.

\section{Results for the general open XXZ spin chain}
\setcounter{equation}{0}

\subsection{Degeneration}

The open boundary XYZ model (\ref{Hamil}) can degenerate to the anisotropic XXZ spin chain with integrable boundary fields.
Taking the trigonometric limit $\tau \rightarrow i \infty$, the couplings constants (\ref{jxyz}) in the Hamiltonian (\ref{Hamil}) degenerates to
\begin{equation}\label{Jxxz}
  J_x = 1, \quad J_y =1, \quad J_z = \cosh(i\pi\eta),
\end{equation}
and the corresponding boundary magnetic fields (\ref{hfx})-(\ref{hfz}) read
\begin{eqnarray}
  h^{\mp}_x &=& \mp i \sinh(i\pi \eta)  \frac{ \sinh(i\pi \beta^\mp_2 )}{  \sinh(i\pi \beta^\mp_1) \cosh(i\pi\beta^\mp_3 ) } ,    \nonumber\\
  h^{\mp}_y &=& \mp  \sinh(i\pi \eta) \frac{\cosh( i\pi\beta^\mp_2 )}{ \sinh(i\pi \beta^\mp_1) \cosh(i\pi \beta^\mp_3 ) } , \nonumber\\
  h^{\mp}_z &=& \mp  \sinh(i\pi\eta)  \frac{\cosh(i\pi \beta^\mp_1 )}{ \sinh(i\pi \beta^\mp_1)} \frac{\sinh(i\pi \beta^\mp_3 )}{\cosh(i\pi \beta^\mp_3 )}. \label{hz-xxz}
\end{eqnarray}
Then the Hamiltonian of open boundary XXZ spin chain can be achieved, and the thermodynamic limit results of open XXZ model can also be obtained by taking a triangular limit of the results of XYZ model.

\subsection{Surface and excitation energies with $|J_z|<1$}

If $\eta$ is real, from Eq.(\ref{Jxxz}) we know that the anisotropic coupling along the $z$-direction of the XXZ spin chain is
characterized by the cosine function, which is smaller than one. After taking the trigonometric limit, the energies (\ref{rbEden})-(\ref{rbEroot}) become
\begin{eqnarray}
&& \bar{e}_{r} =- 2 \frac{\sin(\pi \eta)}{\pi }  \int^{\infty}_{-\infty} \tanh(x \eta)    \frac{\cosh(x (1-2\eta))}{\sinh(x)} d x  - \cos(\pi \eta) , \\
&& \bar{E}^{f}_{r} = - \frac{1}{\pi}\sin(\pi\eta) \int^{\infty}_{-\infty}  \frac{\tanh( \eta x)}{\sinh(x)}  \left\{ \cosh((1-2\eta)x)  + \cosh( (1-|1-2\eta|)x) \right. \nonumber\\
 && \qquad \left. - \cosh( (1-\eta )x)  - \cosh(\eta x)    \right\} d x +  \cos(\pi\eta) - 2  \sin(\pi\eta)\cot(2\pi\eta) ,\\
&& \bar{E}^{\pm}_{r} = - \frac{\sin(\pi\eta)}{\pi}  \int^{\infty}_{-\infty}   \frac{  \tanh( \eta x) }{\sinh( x )}   \left\{ \cosh( (1- 2 \beta^{\pm}_1 )x )  + \cosh( 2 \beta^{\pm}_3 x ) \right\}  d x ,\\
&& \bar{E}^{w}_{r}(w_t) = \left( \sI( w_t +\frac{\eta}{2}i ) - \sI( w_t -\frac{\eta}{2}i )   \right)  \frac{\sin(\pi\eta)}{\pi}  \int^{\infty}_{-\infty }  \frac{ \cosh( 2 w_t i x ) }{ \cosh(\eta x) } d x . \label{trigoEr}
\end{eqnarray}
We note that the interval of the real part of present zero roots is $[-\infty,\infty]$.
From Eq.(\ref{trigoEr}), we find that the discrete roots $\{w_t\}$ located at the infinite boundary contribute nothing to the energy.
From Eqs.(\ref{rbEs}) and (\ref{rsEs}), we obtain the surface energy $\bar{E}^s_r$ of the open boundary XXZ model as
\begin{eqnarray}
  \bar{E}^s_r =  \bar{E}^{f}_r + \bar{E}^{+}_r + \bar{E}^{-}_r + \bar{E}^{w}_r(  \frac{1-\eta }{2}  i ),
\end{eqnarray}
with $\eta\in(\frac{1}{2},1)$ and
\begin{eqnarray}
  \bar{E}^s_r = \bar{E}^{f}_r + \bar{E}^{+}_r + \bar{E}^{-}_r,
\end{eqnarray}
with $\eta\in(0, \frac{1}{2})$.
According to Eqs.(\ref{rbE-ele}) and (\ref{rsE-ele}) with trigonometric limit, the excitation energy $\Delta E_r$ induced by the boundary fields is zero, leading to the result that the excitation is gapless.

As explained previously, we should consider the parameter transformations $\beta^{+}_1 \rightarrow 1- \beta^{+}_1$ and $\beta^{+}_1 \rightarrow  \beta^{+}_1+1$
to cover all the regions of the model parameters. From the analysis in subsection \ref{r-boundary},
it is clear that the change $\beta^{+}_1 \rightarrow 1- \beta^{+}_1$ does not affect the distribution of the discrete roots between two auxiliary lines $\pm \frac{\eta}{2}i$,
while the change $\beta^{+}_1 \rightarrow  \beta^{+}_1+1$ can affect the form of boundary strings.
However, the boundary strings are fixed at the infinity boundary thus it does not have the contribution to the energy.
Then we conclude that the results after taking the transformations $\beta^{+}_1 \rightarrow 1- \beta^{+}_1$ and $\beta^{+}_1 \rightarrow  \beta^{+}_1+1$
are the same as before.

\subsection{Surface and excitation energies with $|J_z|>1$}

If $\eta$ is imaginary, the coupling along the $z$-direction is quantified by the hyperbolic cosine function and the values are larger than one.
After taking the trigonometric limit $\tau \rightarrow\infty i$, Eqs.(\ref{ibEden})-(\ref{ibEroot}) read
\begin{eqnarray}
&& \bar{e}_{i}  = - 4 \sinh(i\pi\eta)  \sum^{\infty}_{k= 1} \tanh (ik\pi \eta) e^{ ik \pi 2\eta}  - \cosh(i\pi\eta) , \\
&& \bar{E}^{f}_{i} =  4\sinh(i\pi\eta) \sum^{\infty}_{k=1} \tanh (i2k\pi \eta)  \left( -e^{4ik\pi \eta} +e^{2ik\pi \eta}   \right) \nonumber\\
 && \qquad  + \cosh(i\pi\eta) -2\sinh(i\pi\eta) \cosh(i\pi\eta) ,\\
&& \bar{E}^{\pm }_{i}  = - 2\sinh(i\pi \eta)  \sum^{\infty}_{k=1 }\tanh (ik\pi \eta) \left( e^{ 2 ik\pi \beta^{\pm }_1}+ e^{ 2ik\pi \beta^{\pm }_3}  \cos(k\pi) \right) ,\\
&& \bar{E}^{w}_{i}(w_t) = - \left(   \sI(w_t + \frac{\eta}{2}) -  \sI(w_t -\frac{\eta}{2}) \right)  \sinh(i\pi\eta)  \sum^{\infty}_{k= -\infty }   \frac{ \cosh( ik\pi 2w_t ) }{\cosh( ik\pi \eta )} .
\end{eqnarray}
The boundary discrete roots associated with (\ref{is-law6-eveng})-(\ref{is-law6-oddg}) at the ground state are
\begin{eqnarray}
  w^{-}_{g} = \frac{1}{2} + \phi_b, \quad  w^{+}_{g} = \frac{1}{2} - \minI \left\{ \frac{\eta}{2} + \beta^{+}_3, 2\eta-\phi_b\right\} , \label{wgp-even}
\end{eqnarray}
with $\phi_b = \minI \left\{ \frac{\eta}{2} + \beta^{-}_3, \eta \right\}$ for the even $N$, and are
\begin{eqnarray}
  w^{-}_{g} = \frac{1}{2} + \maxI \left\{ \frac{\eta}{2} - \beta^{-}_3, 0 \right\}, \quad
  w^{+}_{g} = \frac{1}{2} - \minI \left\{ \frac{\eta}{2} + \beta^{+}_3, \frac{3\eta}{2}+\beta^{-}_3 \right\}, \label{wgp-odd}
\end{eqnarray}
for the odd $N$. Based on them we obtain the surface energy $\bar{E}^s_i$ as
\begin{eqnarray}
  \bar{E}^s_i =  \bar{E}^{f}_{i} + \bar{E}^{+}_{i} + \bar{E}^{-}_{i} +  \bar{E}^{w}_{i}(w^{-}_{g})  + \bar{E}^{w}_{i}(w^{+}_{g}) - \frac{1}{2} (1-(-1)^N) \bar{E}^{w}_{i}(\frac{1}{2}). \label{Es-xxz}
\end{eqnarray}

At the first excited state, the boundary discrete roots associated with (\ref{is-law6-evene})-(\ref{is-law6-odde}) are
\begin{eqnarray}
  w^{-}_{e} = \frac{1}{2} + \maxI \left\{ \frac{\eta}{2}-\beta^{-}_3, 0 \right\} , \quad
  w^{+}_{e} = \frac{1}{2} - \maxI \left\{ \frac{\eta}{2}-\beta^{+}_3,  0 \right\} , \label{wep-even}
\end{eqnarray}
for even $N$, and
\begin{eqnarray}
  w^{-}_{e} = \frac{1}{2} + \frac{\eta}{2} + \beta^{-}_3 , \quad
  w^{+}_{e} = \frac{1}{2} - \maxI \left\{ \frac{\eta}{2}-\beta^{+}_3, 0 \right\} . \label{wep-odd}
\end{eqnarray}
for odd $N$. According to Eqs.(\ref{isE-ele})-(\ref{isEs}), we obtain the excitation energy $\Delta \bar{E}_i$ of the open XXZ model
\begin{eqnarray}
  \Delta \bar{E}_i = \bar{E}^{w}_{i}( w^{-}_{e})  +  \bar{E}^{w}_{i}( w^{+}_{e}) - \bar{E}^{w}_{i}( w^{-}_{g})  -  \bar{E}^{w}_{i}(w^{+}_{g}). \label{Eele-xxz}
\end{eqnarray}

Next, we consider the transformations $\beta^{+}_1 \rightarrow \beta^{+}_1+1 $ and $\beta^{+}_1 \rightarrow - \beta^{+}_1$.
For the XYZ spin chain, the change $\beta^{+}_1 \rightarrow - \beta^{+}_1$ is equivalent to two successive steps changes $\beta^{+}_1 \rightarrow \tau + \beta^{+}_1 \rightarrow \tau- (\tau + \beta^{+}_1)$ deduced by Eq.(\ref{ftrans}).
From the analysis in section \ref{i-boundary}, it is clear that the change $\tau + \beta^{+}_1 \rightarrow \tau- (\tau + \beta^{+}_1)$ do not affect the root patterns.
Thus the change $\beta^{+}_1 \rightarrow - \beta^{+}_1$ is equivalent to the change $\beta^{+}_1 \rightarrow \tau + \beta^{+}_1$.
Therefore, the results of the parameter changes $\beta^{+}_1 \rightarrow \beta^{+}_1+1 $ and $\beta^{+}_1 \rightarrow - \beta^{+}_1$ for the open boundary XXZ spin chain can be directly obtained from the analysis in subsection 4.4.
The change $\beta^{+}_1 \rightarrow \beta^{+}_1+1 $ has no effect on the surface energy $\bar{E}^s_i$ and excitation energy $\Delta \bar{E}_i$.
However, for the change $\beta^{+}_1\rightarrow -\beta^{+}_1$, the energies $\bar{E}^s_i$ and $\Delta \bar{E}_i$ with even (odd) $N$ should be calculated by using the boundary strings with odd (even) $N$, which
depend on the parties of the system.

\section{Conclusions}

In this paper, we study the thermodynamic limit of the XYZ spin chain with the general integrable open boundary conditions.
Although the $U(1)$-symmetry is broken, by using the new parametrization of the eigenvalues of the transfer matrix,
we obtain the surface energy and excitation energy exactly.
We find that the surface energy and excitation energy depend on the parity of the system-size $N$, due to the long-rang Neel order in the bulk.
For the real $\eta$, the system has the spontaneous magnetization and the easy-axis is the $x$-direction.
Flipping the boundary field along the $x$-axis (which occurs in the change $\beta^{+}_1\rightarrow  \beta^{+}_1+1$) has the same boundary and excitation energies as those of changing the parities of $N$.
For the pure imaginary $\eta$, the easy-axis is the $z$-direction.
Flipping the boundary field along the $z$-axis (which occurs in the change $\beta^{+}_1\rightarrow  \beta^{+}_1+\tau$) has the same energy contributions as those of changing the parities of $N$.
We discuss the results for all the regions of model parameters.
We also give the corresponding results for the boundary XXZ model by taking the triangular limit $\tau\rightarrow i\infty$.

\section*{Acknowledgments}

We acknowledge the financial support from National Key R$\&$D Program of China (Grant No.2021YFA1402104),
National Natural Science Foundation of China (Grant Nos. 12074410, 12205235, 12247103, 12247179, 12105003, 11934015 and 11975183), the Major Basic Research Program of Natural Science of Shaanxi Province
(Grant No. 2021JCW-19),  Shaanxi Fundamental Science Research Project for Mathematics and Physics (Grant No. 22JSZ005), and the Strategic Priority Research Program of the Chinese Academy of Sciences (Grant No. XDB33000000).

\section*{Appendix: Elliptic functions}
\label{appA}
\setcounter{equation}{0}
\renewcommand{\theequation}{A.\arabic{equation}}

In this paper, the elliptic functions are defined as
\begin{eqnarray}
  && \theta \left[
           \begin{array}{c}
             a \\
             b \\
           \end{array}
         \right](u,\tau)=\sum_m e^{i \pi (m+a)^2\tau +2i \pi (m+a)(u+b)}, \label{thef}\\
  && \sigma(u) = \theta \left[
           \begin{array}{c}
             \frac{1}{2} \\
             \frac{1}{2} \\
           \end{array}
         \right](u,\tau), \qquad \zeta(u)=\frac{\partial}{\partial u} \{ \ln\sigma(u) \},\label{zetaf}
\end{eqnarray}
where $a$, $b$ are the rational numbers and $\tau$ is the modulus parameter with $\textrm{Im}(\tau)>0$.

The $\sigma$-function satisfies the Riemann-identity
\begin{eqnarray}
  && \sigma(u+x) \sigma(u-x)\sigma(v+y)\sigma(v-y) - \sigma(u+y) \sigma(u-y)\sigma(v+x)\sigma(v-x)  \nonumber\\
  &=& \sigma(u+v) \sigma(u-v)\sigma(x+y)\sigma(x-y).
\end{eqnarray}
Besides, we also use the following identities among the  elliptic functions during the derivation
\begin{eqnarray}
  && \sigma(2u)= \frac{2 \sigma(u)\sigma(u+\frac{1}{2}) \sigma(u+\frac{\tau}{2}) \sigma(u-\frac{1+\tau}{2})     }{ \sigma(\frac{1}{2}) \sigma(\frac{\tau}{2}) \sigma(-\frac{1+\tau}{2}) }  \label{doubleang}\\
  && \sigma(u+1)=-\sigma(u),\qquad \sigma(u+\tau)= - e^{-2i\pi(u+\frac{\tau}{2})} \sigma(u), \\
  && \frac{\sigma(u)}{\sigma(\frac{\tau}{2})} = \frac{ \theta \left[
           \begin{array}{c}
             0 \\
             \frac{1}{2} \\
           \end{array}
         \right](u,2\tau) \theta \left[
           \begin{array}{c}
             \frac{1}{2} \\
             \frac{1}{2} \\
           \end{array}
         \right](u,2\tau) }{ \theta \left[
           \begin{array}{c}
             0 \\
             \frac{1}{2} \\
           \end{array}
         \right](\frac{\tau}{2},2\tau) \theta \left[
           \begin{array}{c}
             \frac{1}{2} \\
             \frac{1}{2} \\
           \end{array}
         \right](\frac{\tau}{2},2\tau)   }, \\
  && \theta \left[
           \begin{array}{c}
             \frac{1}{2} \\
             \frac{1}{2} \\
           \end{array}
         \right](2u,2\tau) =  \theta \left[
           \begin{array}{c}
             \frac{1}{2} \\
             \frac{1}{2} \\
           \end{array}
         \right](\tau,2\tau) \times \frac{\sigma(u)\sigma(u+\frac{1}{2}) }{ \sigma(\frac{\tau}{2})\sigma(\frac{\tau}{2}+\frac{1}{2}) } , \\
  && \theta \left[
           \begin{array}{c}
             0 \\
             \frac{1}{2} \\
           \end{array}
         \right](2u,2\tau) =  \theta \left[
           \begin{array}{c}
             0 \\
             \frac{1}{2} \\
           \end{array}
         \right](0,2\tau) \times \frac{\sigma(u-\frac{\tau}{2})\sigma(u+\frac{1}{2} + \frac{\tau}{2}) }{ \sigma(-\frac{\tau}{2})\sigma(\frac{\tau}{2}+\frac{1}{2}) } .
\end{eqnarray}

\providecommand{\href}[2]{#2}\begingroup\raggedright\endgroup

\end{document}